\documentclass[fleqn,usenatbib]{mnras}

\usepackage{newtxtext,newtxmath}
\usepackage{soul}
\usepackage{caption}
\usepackage{subcaption}
\usepackage[usenames,dvipsnames]{xcolor}

\usepackage[T1]{fontenc}

\DeclareRobustCommand{\VAN}[3]{#2}
\let\VANthebibliography\thebibliography
\def\thebibliography{\DeclareRobustCommand{\VAN}[3]{##3}\VANthebibliography}

\usepackage{graphicx}	
\usepackage{amsmath}	
\usepackage{multirow}

\usepackage{orcidlink}
\usepackage{balance}
\usepackage{booktabs}
\usepackage{siunitx}

\hypersetup{
    colorlinks = true,
    citecolor = {MidnightBlue},
    linkcolor = {BrickRed},
    urlcolor = {BrickRed}
}

\makeatletter
\newcommand*{\rom}[1]{\expandafter\@slowromancap\romannumeral #1@}
\makeatother
\newcommand{\noop}[1]{} 

\title[Disentangling the anisotropic radio sky: Fisher forecasts for 21cm arrays]{Disentangling the anisotropic radio sky: Fisher forecasts for 21cm arrays}

\author[Z. Zhang et al.]{
Zheng Zhang,$^{1}$\,\orcidlink{0000-0002-9154-2803} \thanks{E-mail: zheng.zhang@manchester.ac.uk (ZZ)}
Philip Bull$^{1,2}$\,\orcidlink{0000-0001-5668-3101},
and Katrine A. Glasscock$^{1}$\,\orcidlink{0000-0001-6894-0902}
\\
$^{1}$Jodrell Bank Centre for Astrophysics, University of Manchester, Manchester, M13 9PL, United Kingdom\\
$^{2}$Department of Physics and Astronomy, University of Western Cape, Cape Town 7535, South Africa\\
}

\date{Accepted XXX. Received YYY; in original form ZZZ}

\pubyear{2024}

\begin{document}
\label{firstpage}
\pagerange{\pageref{firstpage}--\pageref{lastpage}}
\maketitle

\begin{abstract}
The existence of a radio synchrotron background (RSB) excess is implied by a number of measurements, including excess emission seen by the ARCADE~2 and LWA experiments. Highly sensitive wideband radio arrays, of the kind used to measure the cosmic 21cm signal, provide a promising way to further constrain the RSB excess through its anisotropy, providing additional insight into its origin. 
We present a framework for evaluating the potential of 21cm arrays to disentangle different components of the diffuse radio sky based on the combination of their frequency spectrum and angular power spectrum (APS).
The formalism is designed to calculate uncertainties due to the intrinsic cosmic variance alone or together with instrumental noise.
In particular, we predict the potential for measuring the anisotropy of a broad generalised class of excess radio background models using the low-frequency HERA array as an example.
We find that a HERA-like array can distinguish an RSB excess from other sky components based on its angular clustering and spectral dependence, even if these are quite similar to one or more of the other components -- but only in the case that the RSB excess is relatively bright.
\end{abstract}




\section{Introduction}
The bright diffuse radio emission at high Galactic latitudes has long been known and has been observed in a number of observations \citep[e.g.][]{westerhout1951comparison}. For a long time, in the absence of good constraints on both the Galactic emission and the line-integrated emission from extragalactic point sources, typical analyses simply assumed that the diffuse radio emission was mainly a combination of Galactic emission and extragalactic point sources. This understanding has led to a lack of motivation to consider any significant radio background, which may be a significant oversight of 21cm cosmology, i.e. the search for the neutral hydrogen lines embedded in the primordial light from Recombination.
Recent re-examinations of the composition of the diffuse radio sky were triggered by ARCADE~2 \citep{fixsen2011arcade}, which used two empirical Galactic foreground models to show that, in addition to the primordial $\sim$3~K blackbody radiation, there is an excess power-law spectrum in the monopole of the diffuse radio background \citep{dowell2018radio},
\begin{equation}
    T_{\rm RSB}(\nu) \simeq 30.4 {\rm K} \left(\frac{\nu}{310 {\rm MHz}}\right)^{-2.66}.
\end{equation}
This excess temperature is also known as the Radio Synchrotron Background (RSB).
Although Synchrotron in this context refers to the power law determined by ARCADE~2, it does not necessarily imply an origin exclusively from synchrotron emission.
Moreover, given the unknown redshifts of the origins of the RSB excess, the terms ``foreground'' and ``background'' are no longer unambiguously defined in 21cm cosmology. To avoid the confusion that these terms may cause, we hereby explicitly divide the radio sky into five segments: Galactic synchrotron, Galactic free-free, Extragalactic free-free, Extragalactic point sources, and RSB excess (excluding, for simplicity, the subdominant 21cm emission).
Note that since extragalactic point sources models or catalogues are usually defined up to a low flux limit, the definition of the RSB excess should not only include unknown exotic contributions (if they exist), but also potential residual extragalactic point sources.

The origin of the RSB excess has become one of the major puzzles in contemporary astrophysics, since both the known Galactic processes and the known classes of extragalactic point sources have difficulty in explaining such a large excess (see \cite{singal2023second} for a recent review). 
A number of exotic models have been proposed. 
Some works have considered the modified population models of faint, unresolved point sources \citep[e.g.][]{vernstrom2014deep, condon2012resolving, hardcastle2021contribution}. 
\cite{biermann2014cosmic} investigated the possibility that supernovae of massive population~\rom{3} stars are the source of the diffuse background.
Some other exotic astrophysical processes have also been considered, such as 
injections from high energy particles \citep{cline2013cosmological},  
emission from stellar black holes \citep{ewall2018modeling}, 
primordial black holes \citep{mittal2022background},
and others more related to structures on cosmological scales, 
such as annihilating dark matter (DM) in halos or filaments \citep{fortes2019some},
superconducting cosmic strings \citep{cyr2023cosmic}.

Strong observational tests are needed for these candidate theoretical models. 
Theoretically, different models have different predictions for { the random field statistics of the RSB excess.
A rich source of information is the frequency-frequency angular power spectrum, which in turn can help to filter out these exotic models}. 
Some recent work has begun to recognise this, e.g. \cite{todarello2023} suggests a 20\% lower bound on the extragalactic contribution to the RSB { (which in this paper should be understood as the sum of the RSB excess, the extragalactic point sources, and the extragalactic free-free emission)} by studying the angular cross-correlation of LOFAR images of the diffuse radio sky with matter tracers at different redshifts provided by galaxy catalogues and CMB lensing.
\citet{offringa2022measurement} and \citet{cowie2023diffuse} investigated the anisotropy of the RSB by examining high multipole modes within fields at high Galactic latitudes. The former analysis attempted to identify a significant Galactic component within the probed range of $\ell$, but concluded that such a component was absent; the latter suggests that while the anisotropy power for $\ell < 7000$ is predominantly due to diffuse Galactic emission, this is not the case for $\ell > 7000$, as indicated by the change in the power law at $\ell \approx 7000$. Nevertheless, there remains a gap in current research methodologies to effectively isolate the RSB excess.  
Dedicated statistical measurements of the frequency and spatial structure of the RSB excess are essential to refine our understanding of the origins of the RSB excess,
especially for an {RSB excess separation} strategy that is logically independent of the ARCADE~2 result and its foreground subtraction strategies.

In this work, to encapsulate our ignorance of each component, we model the diffuse sky as the sum of several independent random fields and constrain them all together. 
We use a generalised parametric form to describe the angular and frequency structures of the random fields. In this way, each field is specified by an independent set of parameters.
The statistical model is described in terms of spherical harmonic (SH) modes whose statistics are characterised by a multivariate Gaussian distribution specified by a parameterised angular power spectrum and frequency-frequency covariance matrix.
Assuming a likelihood-based joint analysis for all fields, we provide a formalism based on Fisher matrix techniques for the prediction of constraints on the RSB excess model using low-frequency radio interferometric arrays. 
The function and purpose of this formalism is twofold: 
(1) to predict whether the intrinsic uncertainty caused by the cosmic variance allows us to disentangle different components, given that we do not have an ensemble of universes to use, but finite SH modes of the field configurations; 
(2) to predict whether the experimental setup allows us to detect the RSB excess and discriminate between different predictions. These are the two things we need to do in order to design an experiment to test particular RSB excess models.

Using this framework, we present a Fisher forecast for the Hydrogen Epoch of Reionization Array \citep[HERA;][]{deboer2017hydrogen} to measure the angular power spectrum (APS) of the RSB excess. Without going into specific theoretical models, we assume three different APS models and analyse the tightness of HERA's constraints on the RSB excess parameters for each of the cases, without imposing any prior on any parameter of any component of the radio sky.

The rest of the paper is structured as follows: 
In Section~\ref{Sec: formalism}, we discuss the basic formalism of Fisher analysis, including a general statistical model (Sec.~\ref{Sec: stat-model}) and the intrinsic uncertainty for constraining the statistical models of the random fields (Sec.~\ref{Sec: cosmic-variance}). 
In Section~\ref{Sec: forecast experiment}, we present a formalism for predicting the ability of the 21cm radio interferometric array to separate different components of the radio sky.
In Section~\ref{sec: setup} we give a specific example where we analyse the ability of the HERA array to detect the RSB excess. 
In Section~\ref{Sec: conclusion} we discuss and conclude this work.

\section{Fisher Forecast Formalism}
\label{Sec: formalism}
Our formalism is based on Fisher matrix techniques and assumes that the diffuse radio sky can be represented as a few statistically independent components, for each of which the distribution of SH modes is approximated by a multivariate complex Gaussian distribution.
This formalism is designed to predict the accuracy of 21cm arrays for disentangling the anisotropic radio sky.
In particular, it can be used to predict their ability to test exotic models for the RSB excess.

\subsection{Statistical model}
\label{Sec: stat-model}
We begin by representing the diffuse radio sky as the sum of several random fields
\begin{equation}
    T(\nu, \theta, \phi) = \sum_{s}  T^{(s)}(\nu, \theta, \phi)\,,
\end{equation}
where the superscript `$s$' indexes the different sky components. 
Typically, our understanding of the sky is in terms of different types of emission sources, such as Galactic synchrotron emission, Galactic free-free emission, one or several RSB excess emissions, etc.

Usually we model the spherical harmonics separately in each small frequency interval, i.e.
\begin{equation}
\label{eq: sky_temperature and Ylm}
    T^{(s)}(\nu, \theta, \phi) = \sum_{\ell,m} a^{(s)}_{\ell m}(\nu) Y_{\ell m}(\theta, \phi),
\end{equation}
where the spherical harmonic coefficient is similarly expressed as the sum of several components
\begin{equation}
    a_{\ell m}(\nu) = \sum_{s}  a^{(s)}_{\ell m}(\nu).
    \label{eq: a_ellm as sum}
\end{equation}
The two-point statistics for each sky component can be characterised by the variance of the spherical harmonic coefficient
\begin{equation}
    \langle a_{\ell m}^{(s)\ast}(\nu_1) a^{(s')}_{\ell' m'}(\nu_2) \rangle
    \equiv
    \delta_{\ell \ell'} \delta_{m m'} \delta_{s s'}C^{(s)}_\ell(\nu_1,\nu_2)\,,
\end{equation}
where $\delta_{s s'}$ constrains that different sky components are statistically independent and $C^{(s)}_\ell(\nu_1,\nu_2)$ is the frequency-frequency angular power spectrum which can be parameterised as \citep{santos2005multifrequency}
\begin{equation}
    C^{(s)}_\ell(\nu_1,\nu_2) = 
    A \, \left(\frac{\ell}{\ell_{\rm ref}}\right)^{ \alpha}
    \left(\frac{\nu_1\nu_2}{\nu_{\rm ref}^2}\right)^{\beta} e^{-\frac{\ln^2\left({\nu_1}/{\nu_2}\right)}{2 \xi^2} },
    \label{eq: aps model}
\end{equation}
where $A$ denotes the magnitude of the power spectrum at the reference frequency $\nu_{\rm ref}$ and the reference scale $\ell_{\rm ref}$. $\alpha$ describes the spatial structure of the random field: the power spectrum of a naturally occurring random field is usually smooth enough that we can approximate it as a power law over a modest interval of $\ell$.
$\beta$ and $\xi$ describe the frequency structure of the radio emission, where $\beta$ describes a power-law approximation of the spectrum and $\xi$ describes the degree of anticorrelation, which can be thought of as a low-order correction to the power law.
We have a set of parameters $\{A^{(s)}, \alpha^{(s)}, \beta^{(s)}, \xi^{(s)}\}$ for each component field indexed by $s$.
The goal of this analysis is to determine whether the values of the parameters $A$, $\alpha$, $\beta$, and $\xi$ for the RSB excess could be constrained by a suitable 21cm interferometer experiment.

\subsection{Cosmic variance and component distinguishability}
\label{Sec: cosmic-variance}
We approximate the distribution of spherical harmonic modes as a multivariate complex normal ($\mathcal{CN}$) distribution. For radio interferometric arrays, if we only consider cross-correlation visibilities, the measurement is basically only sensitive to fluctuating modes and has no response to the monopole. We will therefore only consider the zero mean, $\ell>0$  modes,
\begin{equation}
    \boldsymbol{a} \sim \mathcal{CN}\left(0, \mathbf{C}\right),
\end{equation}
where the sky vector, denoted $\boldsymbol{a}$, is the tuple containing all the spherical harmonic coefficients. 
Each element of the covariance matrix $\mathbf{C}$ is given by the statistical model. The covariance matrix can be represented as the sum over components
\begin{equation}
    \mathbf{C} = \sum_s  \mathbf{C}^{(s)} .
\end{equation}

\begin{table}
    \caption{The settings for four different combinations of spherical harmonic modes.
    To illustrate the effect of cosmic variance and intrinsic model discriminability, each of these sets of $a_{\ell m}(\nu)$ is used to disentangle a four-component radio sky. The results are listed in Table~\ref{tab:cosmic_covariance}. }
\centering
    \begin{tabular*}{\columnwidth}{cccc}
    \toprule
       \multirow{2}{*}{\parbox{1cm}{\centering Forecast \\ name}}        & \multirow{2}{*}{\parbox{2cm}{\centering Frequency \\ range}} & \multirow{2}{*}{\parbox{2cm}{\centering Number of\\ frequencies}} & \multirow{2}{*}{\parbox{2cm}{\centering Range of\\ modes}}     \\ [10pt]
    \midrule
    Low res. & 130-150 MHz     & 5                     & $40\leq\ell\leq 70$\\
    21 freqs. & 130-150 MHz     & 21                    & $20\leq\ell\leq 90$\\
    101 freqs. & 130-150 MHz     & 101                   & $20\leq\ell\leq 90$\\
    Wideband & 130-230 MHz     & 101                   & $20\leq\ell\leq 90$\\
    \bottomrule
    \end{tabular*}
    \label{tab:frequency_SH_combinations}
\end{table}

\begin{figure*}
    \centering
    \includegraphics[width=\textwidth]{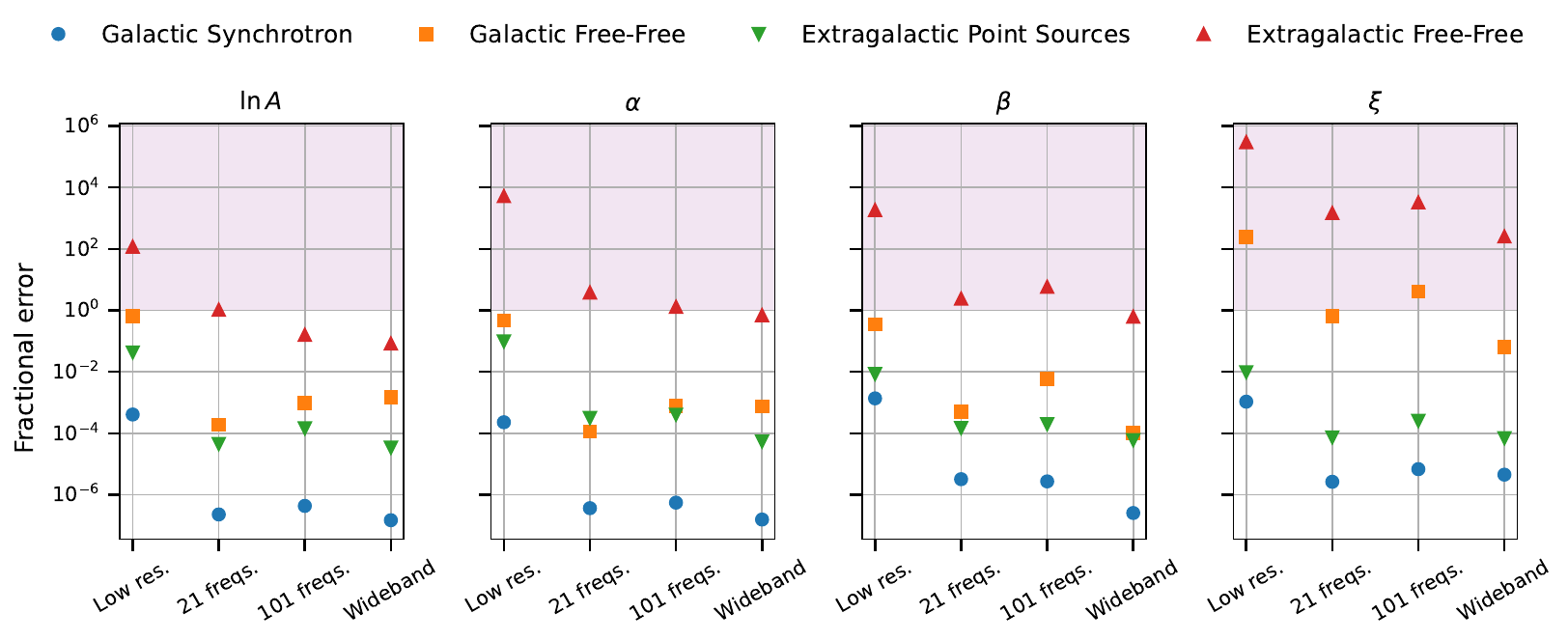}
    \caption{Fractional uncertainty for different parameters due to cosmic variance. The fractional error is obtained by dividing the 1$\sigma$ uncertainty level by the absolute value of the fiducial parameter, i.e. $\sigma/|p_\alpha|$. The shaded areas indicate high fractional uncertainties greater than 1.
    Forecasts \textit{Low res.}, \textit{21 freqs.} and \textit{Wideband}: more spherical harmonic modes at more frequencies provide tighter constraints. 
    Forecasts \textit{21 freqs.}, \textit{101 freqs.} and \textit{Wideband}: a higher frequency resolution doesn't always help.
    In all forecasts, the faint extragalactic free-free component cannot be constrained well.  }
    \label{fig: cosmic-variance}
\end{figure*}

We do not really have an ensemble of universes, but finite spherical harmonics of the field configurations in our Universe. Therefore, before testing the angular power spectrum predicted by the model, we need to know whether the cosmic variance allows us to separate the RSB excess from other components. In other words, the challenge posed by the strategy of constraining all fields simultaneously is whether we can tell the statistical differences between different fields using a finite set of spherical harmonics; if it does not, we are forced to use priors or adapt the experimental setup to look at more frequencies or more spherical harmonic modes.

To characterize the effect of cosmic variance, we assume we have perfectly detected the $a_{\ell m}$'s of the radio sky, i.e. $a_{\ell m}$'s with zero error bar. 
We can calculate the Fisher matrix simply using equation~\eqref{eq: a_ellm as sum} as the data model. The zero mean nature for the $\ell>0$ modes simplifies the Fisher forecasting formalism, as only the second term of the Fisher matrix is retained
\begin{equation}
    \mathcal{F}\left({p_\alpha^{(s)}, p^{(s')}_\beta}\right)  = \frac{1}{2}\mathrm{Tr}\left(\mathbf{C}^{-1}\frac{\partial\mathbf{C}^{(s)}}{\partial p^{(s)}_\alpha}\mathbf{C}^{-1}\frac{\partial\mathbf{C}^{(s')}}{\partial p^{(s')}_\beta}\right),
    \label{eq: cosmic variance fisher}
\end{equation}
where $p_\alpha^{(s)}$ denotes a parameter of the component $s$ and $p^{(s')}_\beta$ denotes a parameter of the component $s'$.
Since each matrix in equation 2 is diagonal with respect to $\ell m$, the Fisher computation can be carried out as a sum over $\ell m$ modes
\begin{equation}
    \mathcal{F}\left({p_\alpha^{(s)}, p^{(s')}_\beta}\right)  
    = 
    \sum_{\ell m}
    \frac{1}{2}\mathrm{Tr}\left(\mathbf{C_\ell}^{-1}\frac{\partial\mathbf{C_\ell}^{(s)}}{\partial p^{(s)}_\alpha}\mathbf{C_\ell}^{-1}\frac{\partial\mathbf{C_\ell}^{(s')}}{\partial p^{(s')}_\beta}\right),
    \label{eq: Fisher formalism for cosmic variance}
\end{equation}
where the square matrix $\mathbf{C_\ell}$ has the size of the number of frequencies.

In order to better discuss the effect of the cosmic variance on the model constraints, we first perform a simple set of comparative analyses.  
We assume a radio sky with four statistically-independent components; two Galactic components (Galactic synchrotron and Galactic free-free), and two RSB components (Extragalactic point sources and Extragalactic free-free).
In presenting an illustrative example of our cosmic variance evaluation scheme, we choose not to include a potential exotic RSB excess. 
This represents the most optimistic case for a statistical model of the anisotropic sky, in which we have a complete model with a reasonably minimal number of statistical parameters. Adding an exotic component (as we shall do in the next section) necessarily increases the number of parameters, and the structure and even the amplitude of the excess is not well constrained.
The fiducial parameter values are given in Section~\ref{sec: setup}. 
We compare how tightly different SH combinations, as listed in Table~\ref{tab:frequency_SH_combinations}, constrain these parameters using equation~\eqref{eq: Fisher formalism for cosmic variance}.
The results of the Fisher forecasts, the simulated measurements and the uncertainties are shown in Table~\ref{tab:cosmic_covariance}, and the uncertainties of each parameter are visualised in Fig.~\ref{fig: cosmic-variance}.
Comparisons between Forecasts \textit{Low res.}, \textit{21 freqs.} and \textit{Wideband} show that, in general, more spherical harmonic modes at more frequencies provide tighter constraints. 
However, comparing Forecasts \textit{21 freqs.}, \textit{101 freqs.} and \textit{Wideband} also shows that it is not always the case that a higher resolution of frequencies is better. 
This can be interpreted as a consequence of the fact that all field components in the model are more or less highly correlated in frequency.
If the difference between two frequencies is much smaller than the correlation length, then the pair will provide no more information than a single frequency. 
Even in extreme cases, { a frequency resolution that is too high} can cause the frequency-frequency covariance matrix to become numerically singular and therefore not accurately inverted.

\begin{table*}
\centering
\renewcommand{\arraystretch}{1.5} 
 \caption{The effect of cosmic variance on the disentanglement of radio sky components, comparison of four different sets of spherical harmonic modes. Table~\ref{tab:frequency_SH_combinations} shows the specifications of each forecast.
 The `Fiducial' column is the input of the Fisher forecasts.
 The forecast results are given as the three numbers, the median and the the upper and lower 1$\sigma$ errors obtained from Markov Chain Monte Carlo (MCMC) simulations. 
 These forecasts do not use any prior or conditionals.
 }
 \label{tab:cosmic_covariance}
 \begin{tabular*}{0.88\textwidth}{ccrcccccccc}
  \toprule
  Components & Parameters & Fiducial & \multicolumn{2}{r}{Low res. }&\multicolumn{2}{r}{21 freqs. }& \multicolumn{2}{r}{101 freqs. }& \multicolumn{2}{r}{Wideband }\\ 
  \midrule
  \multirow{4}{*}{\parbox{2cm}{\centering Galactic \\ Synchrotron}}
  & $\ln{A}$ & $3.79$
  & \multicolumn{2}{r}{  $3.79^{+1.54\times 10^{-3}}_{-1.58\times 10^{-3}}$ }
  & \multicolumn{2}{r}{  $3.79^{+8.60\times 10^{-7}}_{-8.77\times 10^{-7}}$ }
  & \multicolumn{2}{r}{  $3.79^{+1.63\times 10^{-6}}_{-1.64\times 10^{-6}}$ }
  & \multicolumn{2}{r}{  $3.79^{+5.58\times 10^{-7}}_{-5.42\times 10^{-7}}$ }\\ [3.5pt]
  & $\alpha$ & $-2.4$  
  & \multicolumn{2}{r}{  $-2.40^{+5.46\times 10^{-4}}_{-5.53\times 10^{-4}}$ }
  & \multicolumn{2}{r}{  $-2.40^{+8.72\times 10^{-7}}_{-8.56\times 10^{-7}}$ }
  & \multicolumn{2}{r}{  $-2.40^{+1.32\times 10^{-6}}_{-1.29\times 10^{-6}}$ }
  & \multicolumn{2}{r}{  $-2.40^{+3.72\times 10^{-7}}_{-3.75\times 10^{-7}}$ }\\ [3.5pt]
  & $\beta$ & $-2.60$ 
  & \multicolumn{2}{r}{  $-2.60^{+3.54\times 10^{-3}}_{-3.46\times 10^{-3}}$ }
  & \multicolumn{2}{r}{  $-2.60^{+8.29\times 10^{-6}}_{-8.35\times 10^{-6}}$ }
  & \multicolumn{2}{r}{  $-2.60^{+7.03\times 10^{-6}}_{-6.91\times 10^{-6}}$ }
  & \multicolumn{2}{r}{  $-2.60^{+6.57\times 10^{-7}}_{-6.57\times 10^{-7}}$ }\\ [3.5pt]
  & $\xi$ & $4.0$ 
  & \multicolumn{2}{r}{  $4.0^{+4.25\times 10^{-3}}_{-4.34\times 10^{-3}}$ }
  & \multicolumn{2}{r}{  $4.0^{+1.05\times 10^{-5}}_{-1.07\times 10^{-5}}$ }
  & \multicolumn{2}{r}{  $4.0^{+2.71\times 10^{-5}}_{-2.67\times 10^{-5}}$ }
  & \multicolumn{2}{r}{  $4.0^{+1.79\times 10^{-5}}_{-1.87\times 10^{-5}}$ }\\ [3.5pt]
  \midrule
  \multirow{4}{*}{\parbox{2cm}{\centering Galactic \\ Free-Free}}
  & $\ln{A}$ & $-2.43$ 
  & \multicolumn{2}{r}{  $-2.39^{+1.54}_{-1.57}$ }
  & \multicolumn{2}{r}{  $-2.43^{+4.65\times 10^{-4}}_{-4.84\times 10^{-4}}$ }
  & \multicolumn{2}{r}{  $-2.43^{+2.38\times 10^{-3}}_{-2.34\times 10^{-3}}$ }
  & \multicolumn{2}{r}{  $-2.43^{+3.64\times 10^{-3}}_{-3.50\times 10^{-3}}$ }\\ [3.5pt]
  & $\alpha$ & $-3.0$ 
  & \multicolumn{2}{r}{  $-3.03^{+1.41}_{-1.40}$ }
  & \multicolumn{2}{r}{  $-3.00^{+3.44\times 10^{-4}}_{-3.40\times 10^{-4}}$ }
  & \multicolumn{2}{r}{  $-3.00^{+2.29\times 10^{-3}}_{-2.38\times 10^{-3}}$ }
  & \multicolumn{2}{r}{  $-3.00^{+2.23\times 10^{-3}}_{-2.22\times 10^{-3}}$ }\\ [3.5pt]
  & $\beta$ & $-2.15$ 
  & \multicolumn{2}{r}{  $-2.15^{+0.758}_{-0.756}$ }
  & \multicolumn{2}{r}{  $-2.15^{+1.09\times 10^{-3}}_{-1.03\times 10^{-3}}$ }
  & \multicolumn{2}{r}{  $-2.15^{+1.24\times 10^{-2}}_{-1.20\times 10^{-2}}$ }
  & \multicolumn{2}{r}{  $-2.15^{+2.22\times 10^{-4}}_{-2.19\times 10^{-4}}$ }\\ [3.5pt]
  & $\xi$ & $35.0$ 
  & \multicolumn{2}{r}{  $-22.4^{+8.45\times 10^{3}}_{-8.23\times 10^{3}}$ }
  & \multicolumn{2}{r}{  $34.9^{+22.2}_{-21.8}$ }
  & \multicolumn{2}{r}{  $36.6^{+143}_{-149}$ }
  & \multicolumn{2}{r}{  $35.1^{+2.20}_{-2.27}$ }\\ [3.5pt]
  \midrule
  \multirow{4}{*}{\parbox{2cm}{\centering Extragalactic \\ Point Sources}}
  & $\ln{A}$ & $-4.71$ 
  & \multicolumn{2}{r}{  $-4.71^{+0.201}_{-0.198}$ }
  & \multicolumn{2}{r}{  $-4.71^{+2.09\times10^{-4}}_{-2.10\times10^{-4}}$ }
  & \multicolumn{2}{r}{  $-4.71^{+6.75\times10^{-4}}_{-6.61\times10^{-4}}$ }
  & \multicolumn{2}{r}{  $-4.71^{+1.59\times10^{-4}}_{-1.58\times10^{-4}}$ }\\ [3.5pt]
  & $\alpha$ & $-1.1$
  & \multicolumn{2}{r}{  $-1.10^{+0.107}_{-0.105}$ }
  & \multicolumn{2}{r}{  $-1.10^{+3.44\times 10^{-4}}_{-3.48\times 10^{-4}}$ }
  & \multicolumn{2}{r}{  $-1.10^{+4.41\times 10^{-4}}_{-4.40\times 10^{-4}}$ }
  & \multicolumn{2}{r}{  $-1.10^{+5.91\times 10^{-5}}_{-6.09\times 10^{-5}}$ }\\ [3.5pt]
  & $\beta$ & $-2.07$
  & \multicolumn{2}{r}{  $-2.07^{+1.74\times 10^{-2}}_{-1.70\times 10^{-2}}$ }
  & \multicolumn{2}{r}{  $-2.07^{+3.05\times 10^{-4}}_{-3.13\times 10^{-4}}$ }
  & \multicolumn{2}{r}{  $-2.07^{+4.10\times 10^{-4}}_{-4.06\times 10^{-4}}$ }
  & \multicolumn{2}{r}{  $-2.07^{+1.23\times 10^{-4}}_{-1.24\times 10^{-4}}$ }\\ [3.5pt]
  & $\xi$ & $1.0$
  & \multicolumn{2}{r}{  $1.00^{+9.66\times 10^{-3}}_{-9.64\times 10^{-3}}$ }
  & \multicolumn{2}{r}{  $1.00^{+7.31\times 10^{-5}}_{-7.57\times 10^{-5}}$ }
  & \multicolumn{2}{r}{  $1.00^{+2.52\times 10^{-4}}_{-2.48\times 10^{-4}}$ }
  & \multicolumn{2}{r}{  $1.00^{+6.89\times 10^{-5}}_{-6.85\times 10^{-5}}$ }\\ [3.5pt]
  \midrule
  \multirow{4}{*}{\parbox{2cm}{\centering Extragalactic \\ Free-Free}}
  & $\ln{A}$ & $-13.5$ 
  & \multicolumn{2}{r}{  $-8.00^{+1.58\times 10^{3}}_{-1.60\times 10^{3}}$ }
  & \multicolumn{2}{r}{  $-13.5^{+14.1}_{-13.5}$ }
  & \multicolumn{2}{r}{  $-13.5^{+2.14}_{-2.11}$ }
  & \multicolumn{2}{r}{  $-13.5^{+1.12}_{-1.10}$  }\\ [3.5pt]
  & $\alpha$ & $-1.0$ 
  & \multicolumn{2}{r}{  $-32.5^{+5.23\times 10^{3}}_{-5.07\times 10^{3}}$ }
  & \multicolumn{2}{r}{  $-1.01^{+3.80}_{-3.68}$ }
  & \multicolumn{2}{r}{  $-0.0987^{+1.30}_{-1.32}$ }
  & \multicolumn{2}{r}{  $-0.0979^{+0.695}_{-0.706}$ }\\ [3.5pt]
  & $\beta$ & $-2.10$
  & \multicolumn{2}{r}{  $22.6^{+3.87\times 10^{3}}_{-4.00\times 10^{3}}$ }
  & \multicolumn{2}{r}{  $-2.07^{+5.01}_{-5.11}$ }
  & \multicolumn{2}{r}{  $-2.22^{+12.2}_{-12.0}$ }
  & \multicolumn{2}{r}{  $-2.12^{+1.30}_{-1.30}$ }\\ [3.5pt]
  & $\xi$ & $35.0$
  & \multicolumn{2}{r}{  $(1.50\times10^5)^{+1.04\times 10^{7}}_{-1.05\times 10^{7}}$ }
  & \multicolumn{2}{r}{  $-867^{+5.16\times 10^{4}}_{-5.23\times 10^{4}}$ }
  & \multicolumn{2}{r}{  $292^{+1.14\times 10^{5}}_{-1.14\times 10^{5}}$ }
  & \multicolumn{2}{r}{  $235^{+8.91\times 10^{3}}_{-9.28\times 10^{3}}$ }\\ [3.5pt]
  \bottomrule
 \end{tabular*}
\end{table*}

\subsection{Fisher forecast for radio interferometric arrays}
\label{Sec: forecast experiment}
The evaluation of the intrinsic uncertainty due to the cosmic variance is a prerequisite for component separation.
Since the purpose of component separation may vary, the impact of the intrinsic uncertainty caused by the cosmic variance should be evaluated in an ad hoc manner. 
In this analysis we assume that the cosmic variance is not a limit for constraining the RSB excess and other dominant components of the diffuse sky with a particular set of spherical harmonics and frequencies. The next question is whether a 21cm array, in particular its noise, spherical harmonic response and observing strategy, will allow us to separate these components accurately.

The visibility for antennas $i,j$ at frequency $\nu$ and sidereal time $t$ can be written as
\begin{align}
    V(\boldsymbol{b}_{ij}, \nu, t) 
     = \iint \text{d}^2\Omega\, 
    {A_i(\nu, \hat{n}, t)\, A^\dagger_j(\nu,\hat{n}, t)}\,
    {T(\nu, \hat{n})}\, 
    {e^{-2\pi i \nu \tau_{ij}(\hat{n},t)}  },
\end{align}
where $A_{i}$ and $A_j^{\dagger}$ are the E-field beams for each antenna, $\hat{n}$ denotes the direction of the sources and $2\pi\nu\tau_{ij}$ is the phase difference for the source observed by the baseline $\boldsymbol{b}_{ij}$.
Given equation~\ref{eq: sky_temperature and Ylm}, we can rewrite the above using spherical harmonics,
\begin{align}
    V(\boldsymbol{b}_{ij}, \nu, t)  & \equiv \sum_{\ell m} X(\boldsymbol{b}_{ij},\nu, t, \ell, m)\, a_{\ell m}(\nu),
\label{eq: X operator}
\end{align}
where the operator $X$ encodes the response of the array to each spherical harmonic mode on the sky. In linear algebra, $X$ can be represented as a matrix, or a linear mapping projects $\boldsymbol{a}$ into data space as the vector $\boldsymbol{d}$, i.e. $\mathbf{X}: \boldsymbol{a} \longrightarrow \boldsymbol{d}$, where the explicit form is given in Glasscock et al. (in prep). The elements $\delta V_{\ell m}$ of the visibility response operator $\mathbf{X}$ are given as,
\begin{align}
\label{eq:visibility_response_full_definition}
    \delta V_{\ell m}(\boldsymbol{b}_{ij},\nu,t) =  \iint\text{d}^2\Omega\, A_{i}A_{j}^{\dagger}\, Y_{\ell m}(\hat{n})\, e^{-2\pi i \nu \tau_{ij}(\hat{n},t)}.
\end{align}
The visibility response is thus computed for a given array configuration (i.e. given the available baselines, times, and frequencies) per spherical harmonic coefficient $a_{\ell m}$.

Consequently, the statistical model for the data vector is
\begin{equation}
    \boldsymbol{d} \sim \mathcal{CN}\left(0, \mathbf{\Sigma}\right),
\end{equation}
where the covariance matrix is the sum of projected signal and noise according to 
\begin{equation}
    \mathbf{\Sigma} = \sum_s \mathbf{X} \mathbf{C}^{(s)} \mathbf{X}^\dagger + \mathbf{N},
\end{equation}
where $\mathbf{N}$ is the noise covariance matrix.

The Fisher matrix is given by
\begin{equation}
    \mathcal{F}({p_\alpha^{(s)}, p^{(s')}_\beta}) = \frac{1}{2}\mathrm{Tr}\left(\mathbf{\Sigma}^{-1}\frac{\partial\mathbf{\Sigma}}{\partial p^{(s)}_\alpha}\mathbf{\Sigma}^{-1}\frac{\partial\mathbf{\Sigma}}{\partial p^{(s')}_\beta}\right)\,,
    \label{eq: data fisher 1}
\end{equation}
where 
\begin{equation}
    \frac{\partial\mathbf{\Sigma}}{\partial p^{(s)}_\alpha} = 
        \mathbf{X}\frac{\partial \mathbf{C}^{(s)}}{\partial p^{(s)}_\alpha}  \mathbf{X}^\dagger ,
\end{equation}
and $\mathbf{\Sigma}^{-1}$ is usually easier to handle by realising
\begin{equation}
\mathbf{\Sigma}^{-1}=\mathbf{N}^{-1} - \mathbf{N}^{-1} \mathbf{X} (\mathbf{C}^{-1} + \mathbf{X}^\dagger \mathbf{N}^{-1} \mathbf{X})^{-1}  \mathbf{X}^\dagger \mathbf{N}^{-1}\,,
\end{equation}
where we have applied the Woodbury matrix identity.
After some algebra steps, equation~\eqref{eq: data fisher 1} can be rewritten as
\begin{equation}
    \mathcal{F}({p_\alpha^{(s)}, p^{(s')}_\beta}) = \frac{1}{2}\mathrm{Tr}\left(
    \mathbf{M}\frac{\partial\mathbf{C}^{(s)}}{\partial p^{(s)}_\alpha}
    \mathbf{M}\frac{\partial\mathbf{C}^{(s')}}{\partial p^{(s')}_\beta}
    \right),
    \label{eq: data fisher 2}
\end{equation}
where 
\begin{equation}
    \mathbf{M}= \mathbf{Q} - \mathbf{Q} (\mathbf{C}^{-1}+\mathbf{Q})^{-1} \mathbf{Q}\,,
\end{equation}
and $\mathbf{Q}=\mathbf{X}^\dagger \mathbf{N}^{-1} \mathbf{X}$.

\subsubsection{Universal SED approximation}
\label{sec: universal SED}
The representation of diffuse emission using spherical harmonic coefficients defined for each frequency interval is a common way of modelling it and has the advantage that the formalism looks neat. 
However, it also means that we are dealing with a particularly large total number of spherical harmonics (number of frequencies multiplied by number of spherical harmonic modes). 
If such large matrices cannot be handled numerically, the following universal spectral energy distribution (SED) approximation can be considered. This should work well if the field of view is small and the antenna side lobes are well apodised.

More specifically, the simplified toy model for each component would be a universal power-law SED multiplying a mildly anisotropic temperature distribution on the sky, i.e.
\begin{equation}
    T^{(s)}(\nu, \theta, \phi) = \sum_{\ell,m} a^{(s)}_{\ell m} Y_{\ell m}(\theta, \phi) f^{(s)}(\nu),
\end{equation}
where the SED is assumed to have a power-law like structure, 
\begin{equation}
    f(\nu) =\left(\frac{\nu}{\nu_{\rm ref}}\right)^{\beta\left(\ln{\frac{\nu}{\nu_{\rm ref}}}\right)},
\end{equation}
and we have expressed the power index as a frequency-dependent function $\beta$, to inscribe any deviation from the power-law.
On a log-log scale, this SED function can be rewritten as
\begin{equation}
    \ln f(\nu) = {x\beta(x)}
    = {\beta(0) x + \beta'(0)x^2 + \frac{\beta''(0)}{2} x^3 + \cdots },
\end{equation}
where $x\equiv \ln{(\nu/\nu_{\rm ref})}$; In the second equality, we have Taylor expanded $\beta$ at $\nu_{\rm ref}$.
    
The frequency covariance part of $C_\ell(\nu_1,\nu_2)$ can be rewritten in $x$ coordinates as
\begin{equation}
\exp{\left[\beta (x_1+x_2) - \frac{(x_1 - x_2)^2}{2\xi^2}\right]}.
\end{equation}
When characterizing the same diffuse emission using these two models, we have
\begin{equation}
\sum_{n=0}^{n_c} \frac{\beta_{n}}{n!} (x_1^{n+1}+x_2^{n+1}) = 
\beta (x_1+x_2) - \frac{(x_1 - x_2)^2}{2\xi^2},
\label{eq: fitting universal SED model}
\end{equation}
where $\beta_{n}\equiv d^n \beta/dx^n$ can be found by fitting the left to the right and $n_c$ is the cut-off order of the Taylor expansion.

Using equation~\eqref{eq: fitting universal SED model}, we can build statistical models of the parameters $\{A^{(s)}, \alpha^{(s)}, \beta_0^{(s)}, \dots, \beta_n^{(s)}\}$, which is equivalent to the equation~\eqref{eq: aps model} under the universal SED approximation, but with a significantly reduced size of sky modes, denoted as the reduced sky vector $\boldsymbol{a'}$. The mapping between $\boldsymbol{a'}$ and the data vector $\boldsymbol{d}$ is given by
\begin{equation}
    V(\boldsymbol{b}_{ij}, \nu, t) 
     =
     \sum_{\ell m} X(\nu, t, \ell, m) f^{(s)}(\nu) a^{(s)}_{\ell m}.
\end{equation}
Or in the linear algebra form, we can define $\mathbf{X'}: \boldsymbol{a'} \longrightarrow \boldsymbol{d}$, and the matrix $\mathbf{X'}$ is described by
\begin{equation}
    \mathbf{X'}
    \equiv
    \begin{pmatrix}
         f^{(1)}(\nu_0) \mathbf{X}(\nu_0)  & \dots & f^{(s)}(\nu_0) \mathbf{X}(\nu_0)\\
        \vdots  &  \ddots &  \vdots  \\
         f^{(1)}(\nu_N) \mathbf{X}(\nu_N)  & \dots & f^{(s)}(\nu_N) \mathbf{X}(\nu_N)\\
    \end{pmatrix} .
\end{equation}
The covariance matrix of $\boldsymbol{d}$ is then given by 
\begin{equation}
    \mathbf{\Sigma} =  \mathbf{X'} \mathbf{C'} \mathbf{X'}^\dagger + \mathbf{N}.
\end{equation}
To calculate the Fisher matrix (see Eq.~\ref{eq: data fisher 1}), we also need the derivatives of $\mathbf{\Sigma}$, which is quite direct: 
For $p_\alpha = A^{(s)}, \alpha^{(s)}$, we have
    \begin{equation}
        \frac{\partial\mathbf{\Sigma}}{\partial p_\alpha} = 
        \mathbf{X'}\frac{\partial \mathbf{C'}}{\partial p_\alpha} \mathbf{X'^\dagger}, 
    \end{equation}
while for $p_\alpha = h_0^{(s)}, \dots, h_n^{(s)}$, it is
    \begin{equation}
        \frac{\partial\mathbf{\Sigma}}{\partial p_\alpha} = \frac{\partial \mathbf{X'}}{\partial p_\alpha}  \mathbf{C' X'}^\dagger +  \mathbf{X'C'}\frac{\partial  \mathbf{X'}^\dagger}{\partial p_\alpha} .
    \end{equation}

\section{Fisher Forecast for HERA}
\label{sec: setup}

\begin{figure}
    \centering
    \includegraphics[width=\columnwidth]{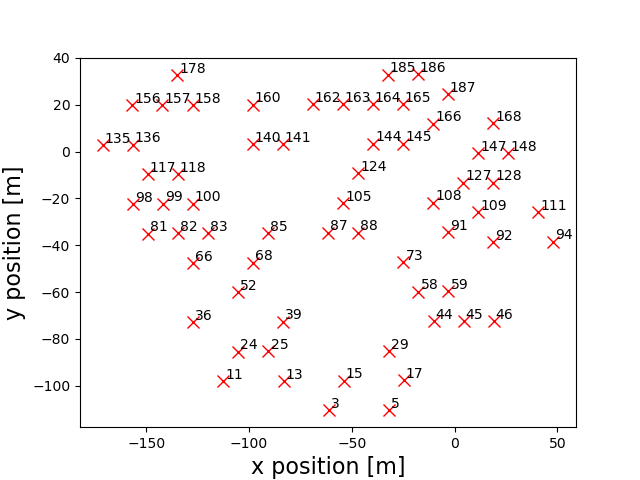}
    \caption{The layout of the reduced dish array.}
    \label{fig:array-layout}
\end{figure}
In this section we analyse the ability of the HERA array to separate different radio components. We briefly present the experimental setup, the fiducial models for the Fisher forecast, and the results.

\subsection{Experimental Setup}
The HERA 21cm array is composed of 14~m parabolic dishes with wide-band Vivaldi feeds suspended at prime focus. We simulate HERA Band 1 (roughly from 115~MHz to 135~MHz) observations in a relatively radio quiet part of the survey region: Field C (RA: 4.0--6.25 hours).

We use the \texttt{hydra}\footnote{\url{https://github.com/HydraRadio/}} diffuse sky sampler 
(Glasscock et al., in prep), which uses the \texttt{healpix} scheme and a \texttt{matvis}-based visibility simulator \citep{kittiwisit2023matvis}, to generate the per-baseline visibility response to the unit sky SH mode, i.e. the $\mathbf{X}$ operator in equation~\eqref{eq: X operator}. 
We use this to test a particular RSB excess theory for a range of critical angular scales of $\ell\leq 90$, specifically we choose $\ell_{\rm max}=90$ and $N_{\rm side}=64$. We approximate the beams as Gaussians where the FWHM is given as $\sim\lambda/D$, where $D$ is the diameter of the dishes. For our frequency range this corresponds to a ${\rm FWHM}$ from $\sim\SI{10.6}{\degree}$ to $\sim\SI{9.09}{\degree}$.

\subsubsection{Baseline selection}
In order to reduce unnecessary numerical work due to duplicate baselines, we use a subset of the actual HERA array layout, as shown in Fig.~\ref{fig:array-layout}.
We then manually re-weight the baseline densities to solve the problem that the population of short baselines in this simplified array layout is significantly smaller than the real number.
Since the chosen range of the targeted and thus modelled SH modes is $\ell \leq 90$,
we impose a baseline filter to avoid the energy of larger-$\ell$ SH modes. 
A rough estimate is that a baseline of length $d$ is sensitive to SH modes $\ell\sim{\pi d}/{\lambda}$.
Therefore, the filter removes baselines longer than $\sim 75$ metres. 
Note that unlike Sec.~\ref{Sec: forecast experiment} we do not set a cutoff $\ell_{\rm min}$ although the minimum baseline length ($\SI{14.6}{\meter}$) sets an effective minimum $\ell$ mode instead.

\subsubsection{Noise}

For the visibility measurements at each time and frequency, we assume a Gaussian random noise uncorrelated with the signal components, $n_{ij}(\nu, t) = \sigma_{ij}({a+ib})/{2}$.
The noise variance is modelled using the simulated autocorrelation visibilities as follows 
\begin{equation}
    \sigma_{ij}^2 = \frac{V_{ii} V_{jj}}{N_{\rm days} \Delta t \Delta\nu},
\end{equation}
where $\Delta\nu = 166$~kHz is the frequency bandwidth and $\Delta t = 40$~sec is the time resolution. $N_{\rm days} = 40$ is the assumed total number of days of observation. The autocorrelation visibilities are generated using an improved model of diffuse Galactic radio emission \citep{de2008model}.

\subsection{Fiducial models}

\begin{table}
    \caption{Fiducial parameter values at $\ell = 10$ and $130$ MHz. 
    These values are the input for the Fisher forecast in Sec.~\ref{sec: setup}.}
\centering
    \begin{tabular*}{\columnwidth}{ccccc}
    \toprule
       \multirow{2}{*}{\parbox{1.5cm}{}}        
       & \multirow{2}{*}{\parbox{1cm}{\centering $A$ }} 
        & \multirow{2}{*}{\parbox{1cm}{\centering   $\alpha$}} 
        & \multirow{2}{*}{\parbox{1cm}{\centering  $\beta$}} 
        & \multirow{2}{*}{\parbox{1cm}{\centering  $\xi$}}   \\ [10pt]
    \midrule
    Gal. synch.& $(6.63 \text{ K})^2$         & $-2.4$          &  $-2.60$   &  $4.0$\\
    Gal. FF.& $(0.30 \text{ K})^2$        & $-3.0$          &  $-2.15$   &  $35.0$\\
    Extragal. ptsrc.& $(0.095\text{ K})^2$  & $-1.1$          &  $-2.07$   &  $1.0$\\
    Extragal. FF.& $(0.0012\text{ K})^2$ & $-1.0$          &  $-2.10$   &  $35.0$\\
    \midrule
    RSB excess 1  & $(6.63 \text{ K})^2$         & $-2.4$          &  $-2.66$   &  $4.0$\\
    RSB excess 2  & $(2.10 \text{ K})^2$         & $-3.0$          &  $-2.66$   &  $4.0$\\
    RSB excess 3  & $(21\text{ K})^2$         & $0.0$           &  $-2.66$   &  $4.0$\\
    \bottomrule
    \end{tabular*}
    \label{tab:fiducial_parameters}
\end{table}

\begin{figure}
    \centering
    \includegraphics[width=\columnwidth]{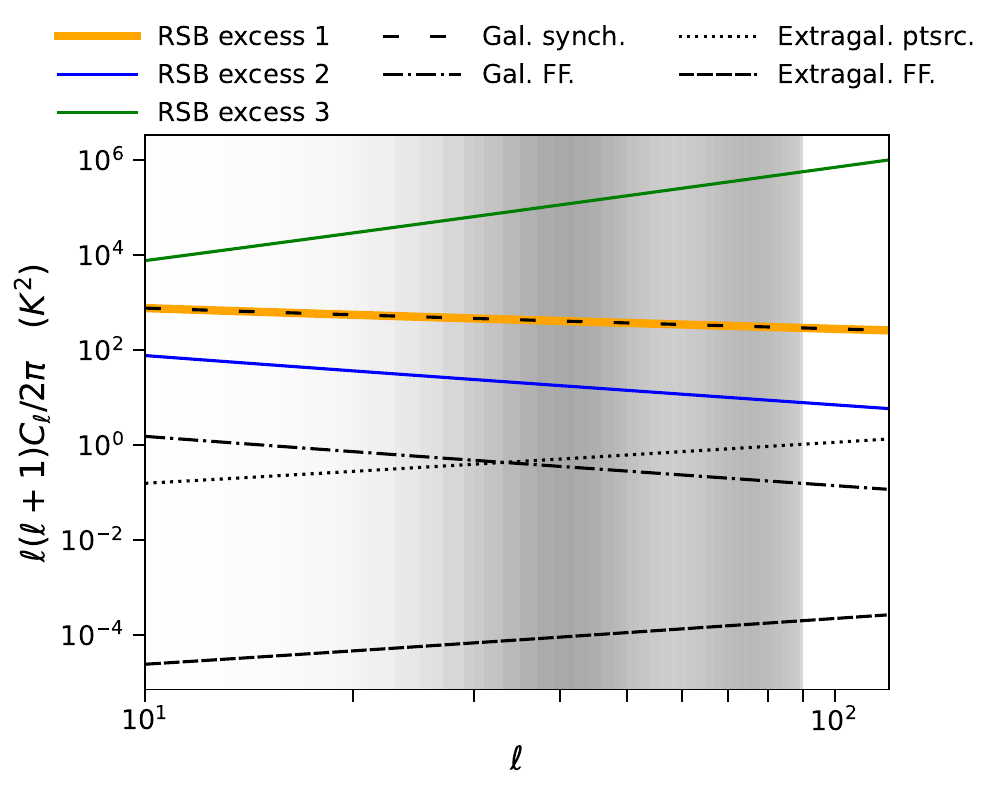}
    \caption{The assumed angular power spectra of different components of the radio sky at the reference frequency of 130 MHz. 
    Note that the RSB excess 1 overlaps the Galactic synchrotron component.
    The opacity of the shaded background is proportional to the overall response of the HERA array model used in this work to $C_\ell$. 
    The overall $C_\ell$ response, which is defined as $\sum_{t, m, \boldsymbol{b}_{ij}}|X(\boldsymbol{b}_{\rm ij},\nu_{\rm ref}, t, \ell, m)|^2/\sigma_{\rm ij}^2(t)$, characterizes the linear mapping from $C_\ell$ to the whitened data covariance, $\langle\boldsymbol{d}^\dagger \mathbf{N}^{-1} \boldsymbol{d} \rangle$.
    For computational reasons we have only calculated up to $\ell_{\rm max}=90$. 
    }
    \label{fig:angular ps}
\end{figure}

We assume that the radio sky consists mainly of five statistically independent components, namely Galactic synchrotron, Galactic free-free, extragalactic point sources, extragalactic free-free and the radio synchrotron background.
For the first four, 
we use Eq.~\ref{eq: aps model} and the parameter values that were presented in \citet{santos2005multifrequency} as the fiducial values for the Fisher analysis (see Table~\ref{tab:fiducial_parameters} and Fig.~\ref{fig:angular ps}),
{
except for the Galactic synchrotron spectral index, for which we use $\beta=-2.6$. This is a typical value for high Galactic latitudes between 90 and 190 MHz \citep{2021Padovani}.
This fiducial Galactic synchrotron power law is flatter than the fiducial RSB power law, reflecting the current perception in the RSB literature \citep[e.g.][]{dowell2018radio} that the RSB excess plays a more important role at lower frequencies.
Note also the chosen spectral index of $-2.07$ for the extragalactic point sources used in \cite{santos2005multifrequency}. This flattened spectrum was obtained by first extrapolating the spectral index in the low frequency interval down to 0.1~mJy, the low flux cutoff below which the clustering signal dominates over the Poisson signal. It was then extrapolated to the spectral index in the lower frequency interval based on the observation that the source spectra flattened from the `low' frequency interval of 327 MHz to 1.4 GHz to the `very low' frequency interval of 74 MHz to 327 MHz, with a median change in the spectral index of $\Delta\beta = 0.24$. 
}

For the RSB excess, since there are currently no well-motivated fiducial values other than the frequency power law index, i.e. $\beta\simeq -2.66$, 
we consider three simple phenomenological models, using the Galactic Synchrotron as a reference, to test the ability of HERA to constrain the RSB excess parameters under different circumstances:
\begin{itemize}
    \item \textbf{RSB {excess 1}}: {The parameters, except for $\beta$, are the same as for Galactic synchrotron. This model mimics the Galactic synchrotron in terms of its angular power spectrum and amplitude, differing from it only in its frequency structure. We include this as a toy model for an RSB excess that would be difficult to distinguish from foreground emission based on angular clustering statistics alone.}

    \item \textbf{RSB excess 2}: 
    { At the reference point ($\ell_{\rm ref}=10, \nu_{\rm ref}=130 \text{ MHz}$), the anisotropy power of the RSB is an order of magnitude less than the Galactic synchrotron, but has a steeper angular power law.}

    \item \textbf{RSB excess 3}:
    { At the reference point ($\ell_{\rm ref}=10, \nu_{\rm ref}=130 \text{ MHz}$), the anisotropy power of the RSB is an order of magnitude stronger than that of the Galactic synchrotron, but has a flatter angular power law.}
\end{itemize}
{
The RSB excess 1 is not physically motivated and we discuss the discriminability of the model more in a numerical sense, as proximity in parameter values could lead to degeneracies in the model, where different combinations of parameter values could lead to similar predictions.  
On the other hand, RSB excess 2 has a steeper APS where the larger (angular) scale fluctuations dominate. 
RSB excess 3 is an example of a flatter APS, where smaller (angular) scale fluctuations play a more important role. This particular model represents white noise extragalactic emission. All three of these cases are intended to be toy models that represent phenomenological scenarios, and are not intended to mimic particular physical explanations for the excess.
}
Theoretically, we could have also considered a higher (weaker) power and a steeper (flatter) angular power law. However, such an assumption implies that the monopole {of the RSB excess} is much stronger (weaker) than the Galactic synchrotron monopole, considering a brutal extrapolation of the APS model to the $\ell \to 0$ end. Such 
low $\ell$ structures do not seem to be supported by the current monopole measurements. 
Also, as an illustrative example, we do not take the trouble to exhaust all the other possible models.

\subsection{Results}
\label{sec: hera results}

\begin{figure*}
     \centering
     \begin{subfigure}[b]{0.44\textwidth}
         \centering
         \includegraphics[width=\textwidth]{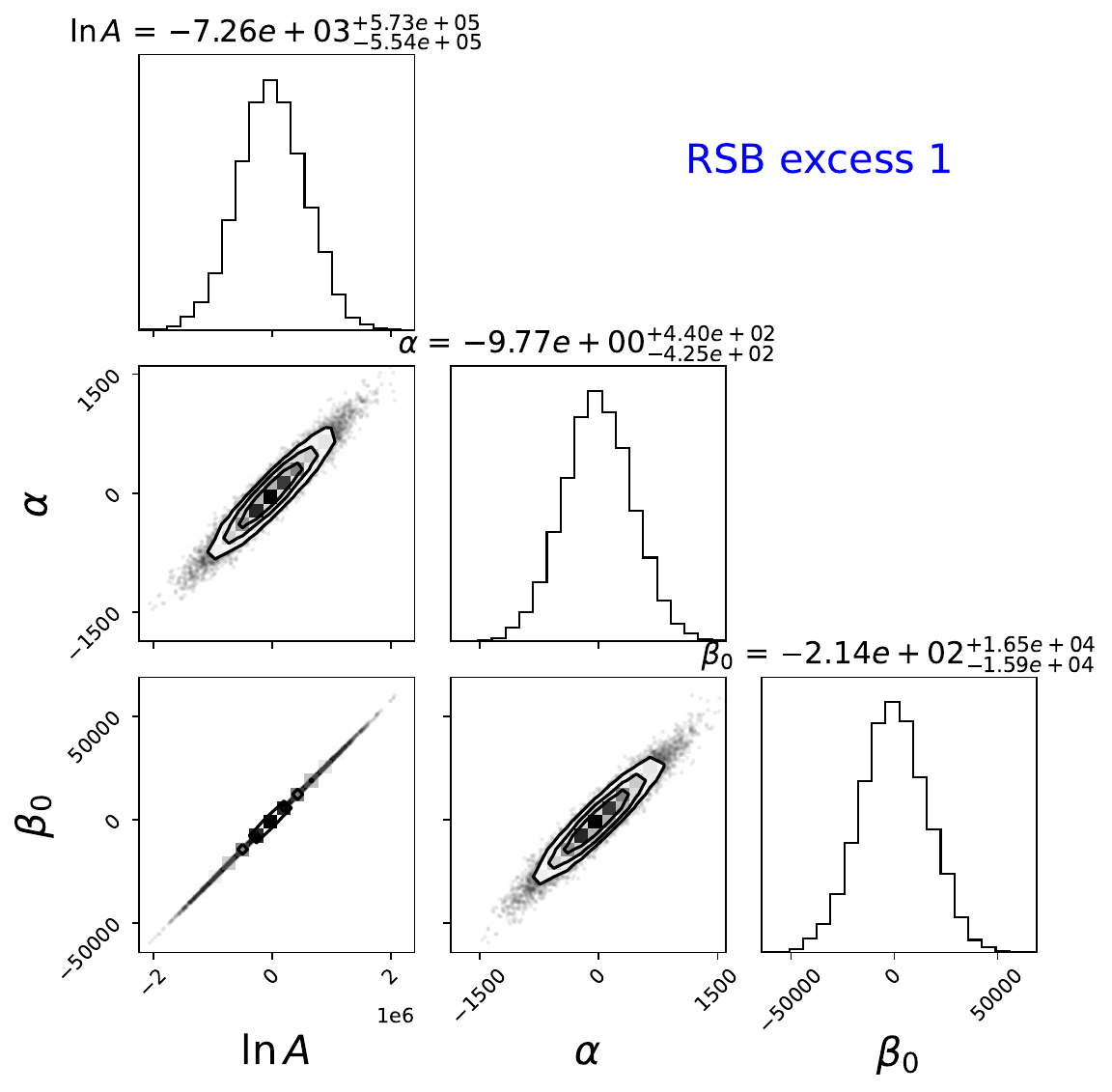}
     \end{subfigure}
     \hfill
     \begin{subfigure}[b]{0.44\textwidth}
         \centering
         \includegraphics[width=\textwidth]{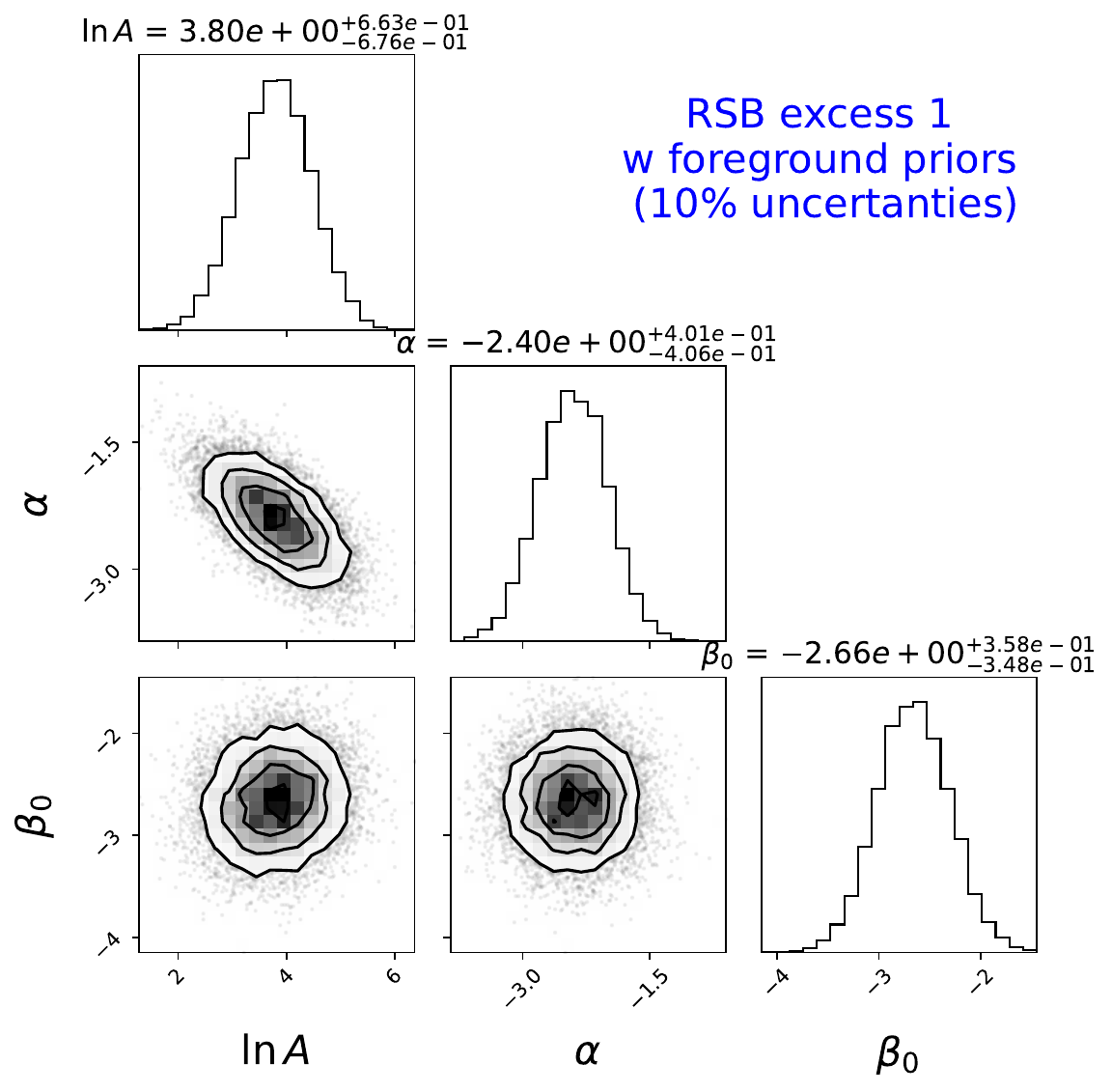}
     \end{subfigure}
     \hfill
     \begin{subfigure}[b]{0.44\textwidth}
         \centering
         \includegraphics[width=\textwidth]{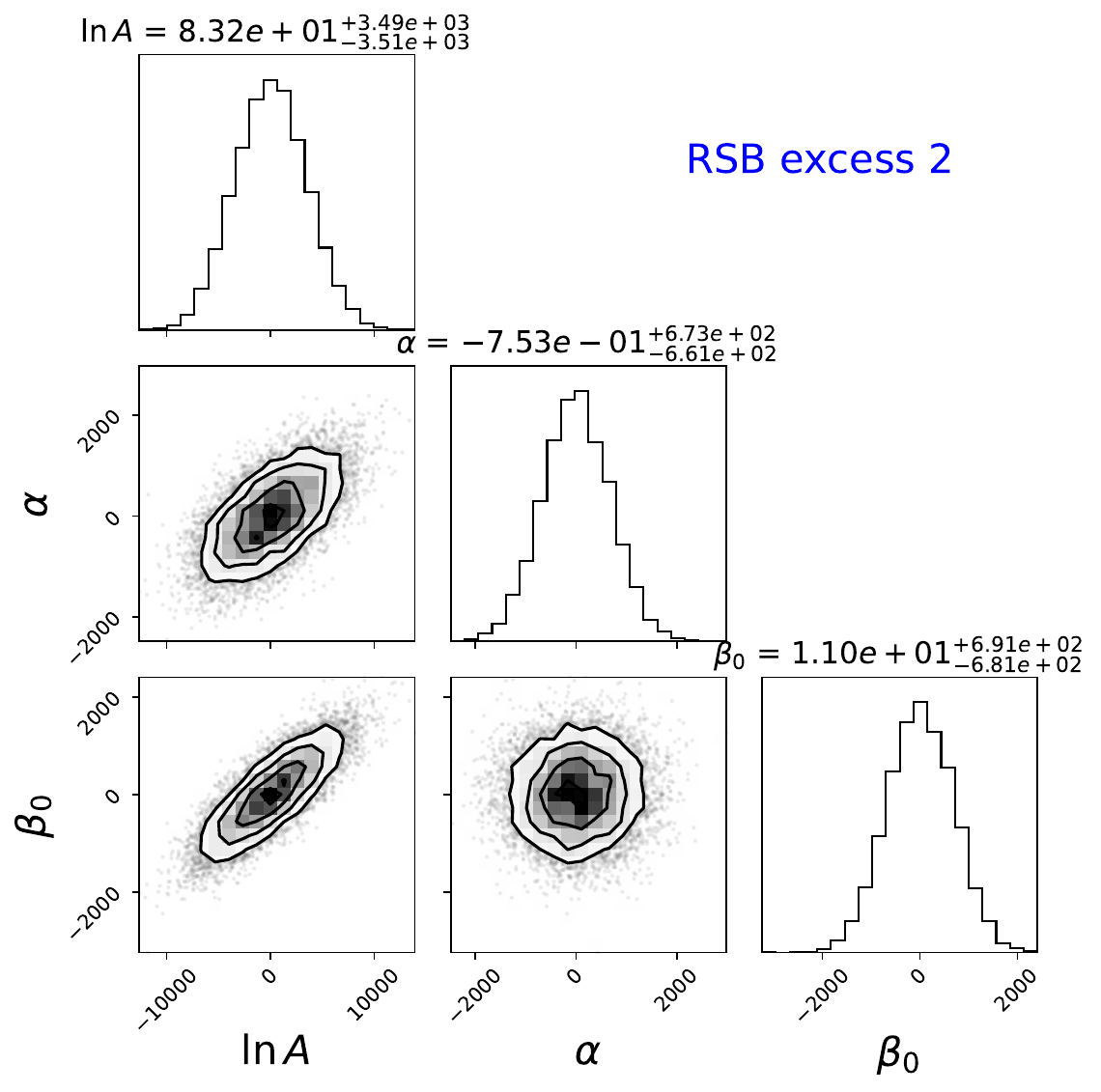}
     \end{subfigure}
     \hfill
    \begin{subfigure}[b]{0.44\textwidth}
         \centering
         \includegraphics[width=\textwidth]{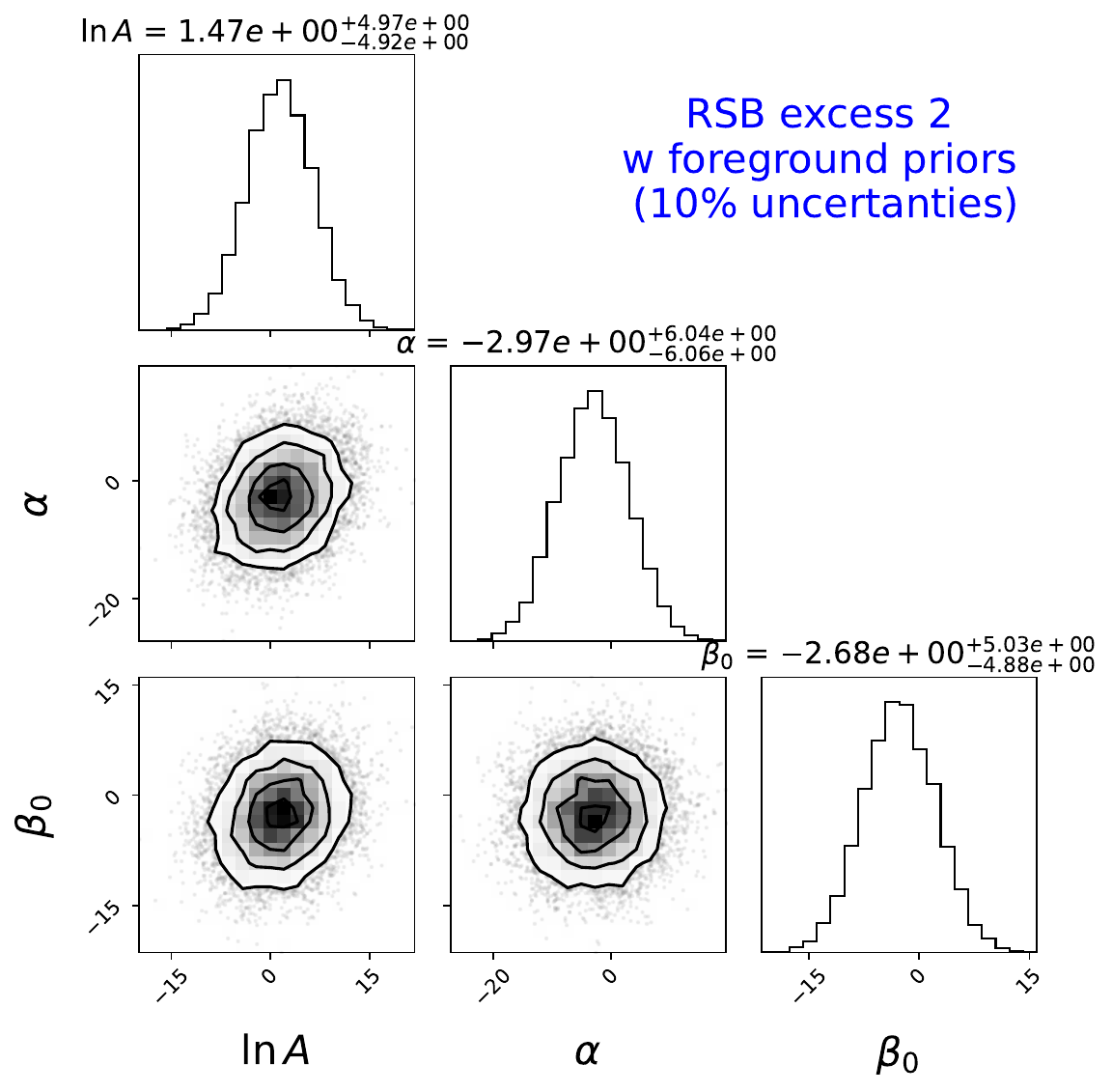}
     \end{subfigure}
     \hfill 
     \begin{subfigure}[b]{0.44\textwidth}
         \centering
         \includegraphics[width=\textwidth]{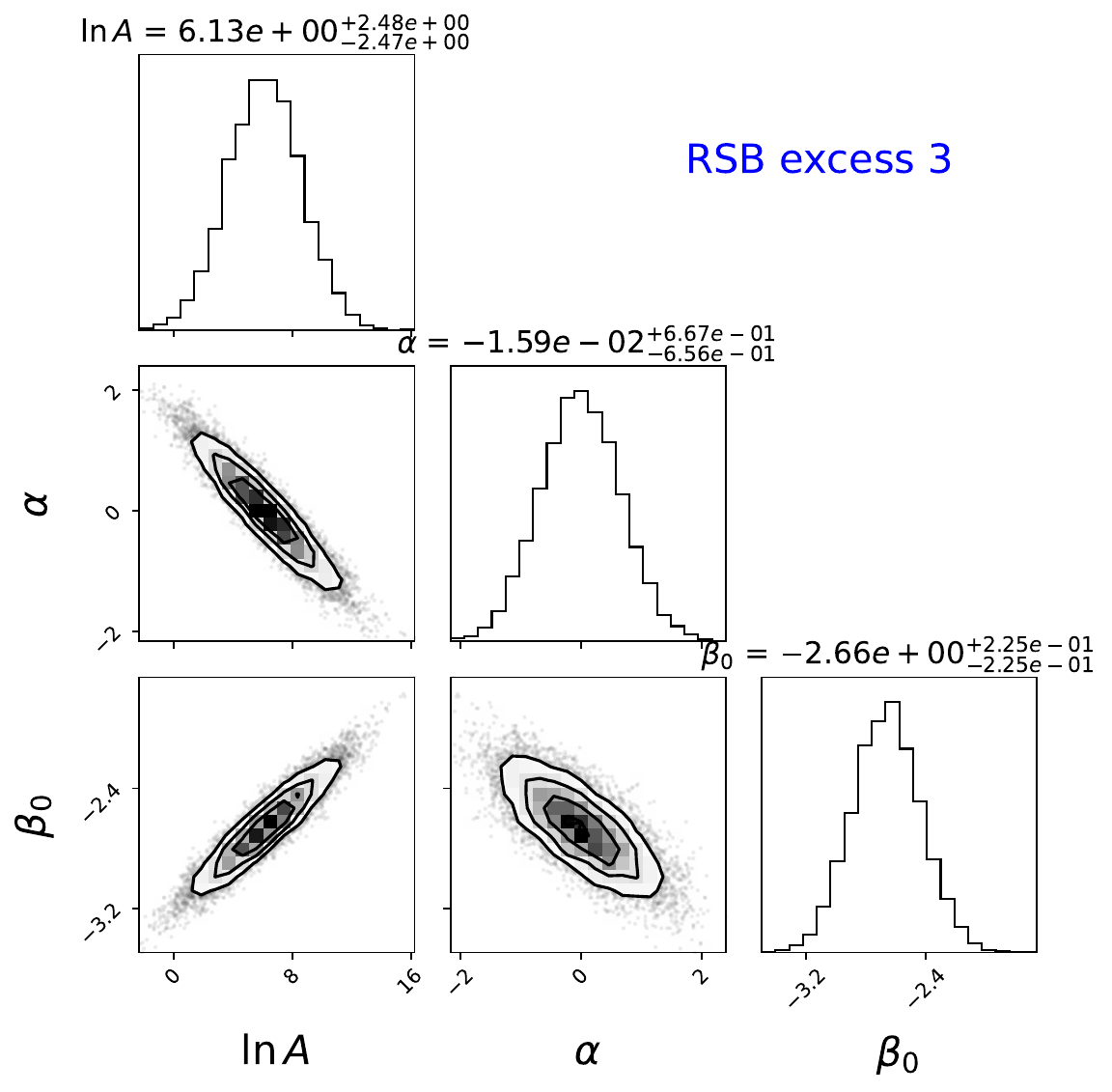}
     \end{subfigure}
    \hfill
    \begin{subfigure}[b]{0.44\textwidth}
         \centering
         \includegraphics[width=\textwidth]{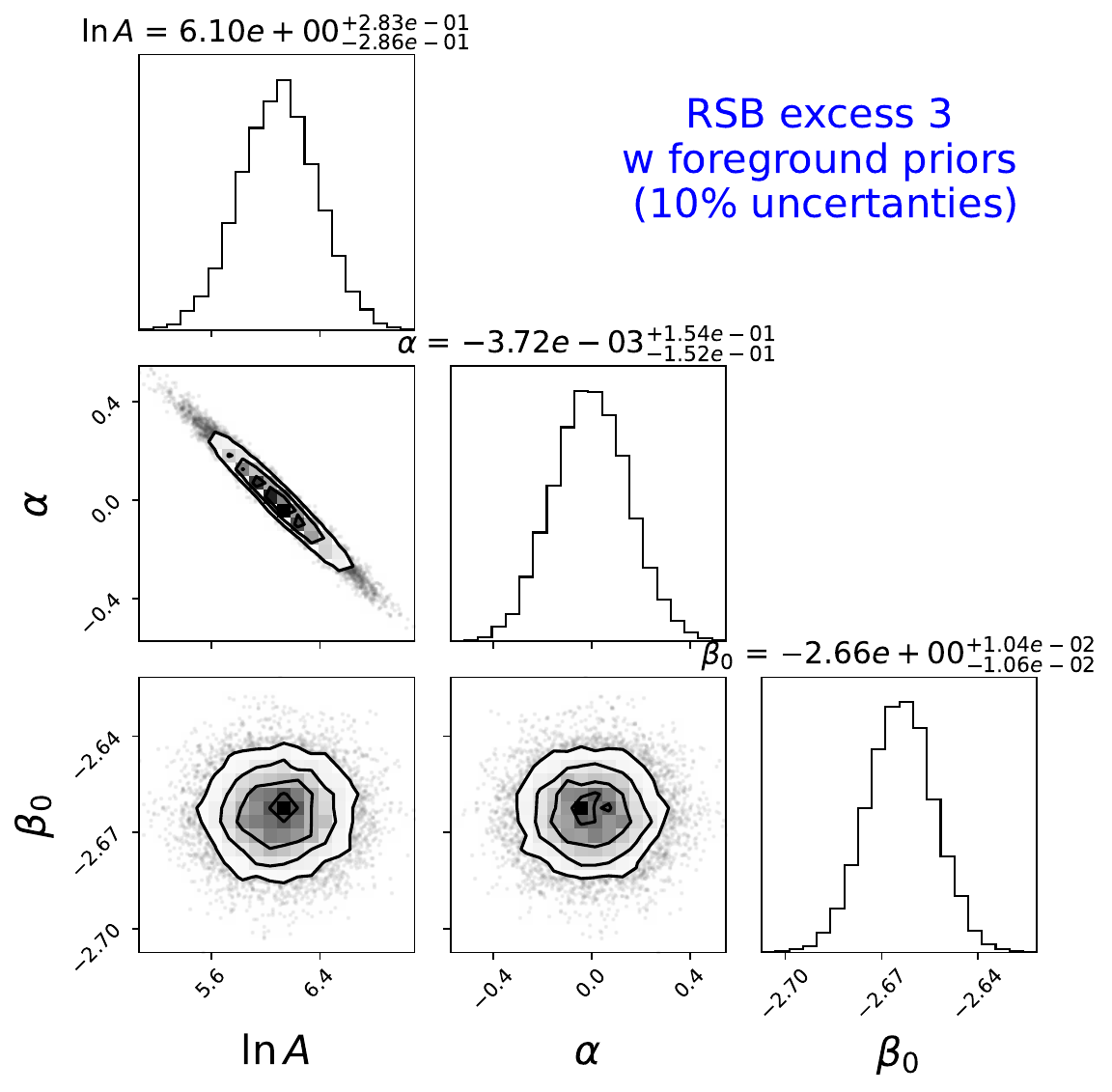}
     \end{subfigure}
        \caption{Fisher forecasts for different RSB models. The left column shows the results with flat priors; the right column shows the results assuming that the foreground parameters are known with an uncertainty of 10\%, but flat priors for the RSB parameters.
        }
        \label{fig:RSB-corner-plots}
\end{figure*}

\begin{figure*}
    \centering
    \includegraphics[width=\textwidth]{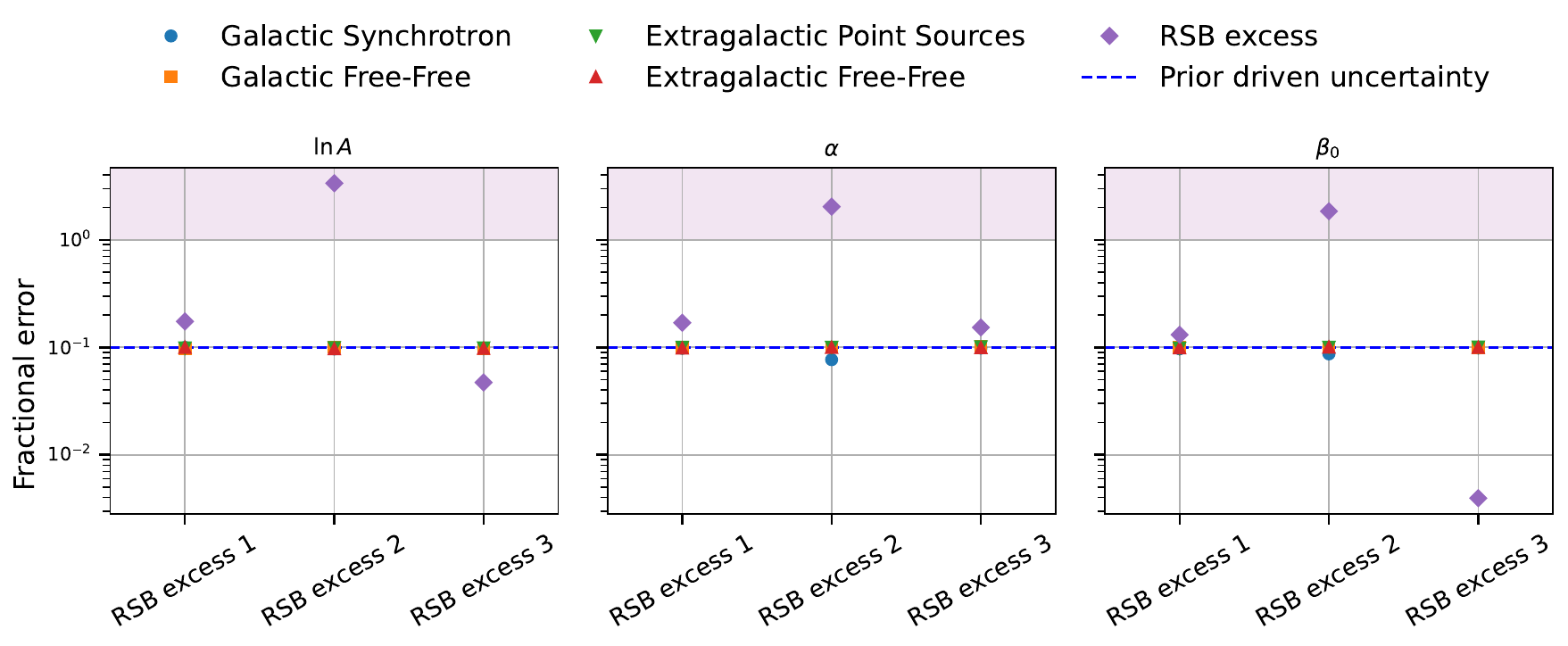}
    \caption{Forecast fractional uncertainty on diffuse sky parameters from HERA, using the 10\% uncertainty foreground priors. The $x$ axis indexes forecasts considering different RSB excess models. It shows that the foreground parameter uncertainties are significantly constrained by the priors.
    It can be seen that RSB excess 1 and 3 are tightly constrained with a subdominant level uncertainty, while RSB excess 2 is far from well constrained in this experimental setup. 
    The fractional error is evaluated as 1$\sigma$ divided by the absolute value of the fiducial parameter, or just $1\sigma$ if the fiducial value is 0.
    The shaded areas indicate high fractional uncertainties greater than 1.}
    \label{fig:hera-uncertainties}
\end{figure*}

We assume three different radio skies, with the respective RSB excess components described by RSB excess 1, 2 and 3, while all three cases are identical in the remaining four components. 
We investigate the ability of HERA to constrain these RSB excess parameters in two scenarios: (1) without using any priors; (2) using priors on the four foreground components (i.e. 10\% uncertainties on each foreground parameter).

Due to the numerical challenges of inverting large covariance matrices when computing Fisher matrices,
we use the universal SED approximation (up to $0$th order) to reduce the matrix size, as explained in Sec.~\ref{sec: universal SED}. 
In this toy model analysis, we only consider the frequency structure up to the zeroth order, i.e. the power law.
In the case of no prior on any parameter of any component, the result of the Fisher analysis shows that the RSB excess model can hardly be constrained with the experimental setup in this work, as shown in Fig.~\ref{fig:RSB-corner-plots}.
On the other hand, the right column of the same figure, as well as Fig.~\ref{fig:hera-uncertainties}, shows that when we apply the priors to the uncertainties of the foreground parameters, the constraints on the RSB excess parameters become tighter. In particular, RSB excess 1 and 3 are constrained with a sub-dominant level of uncertainty, while RSB excess 2 is still far from being well constrained.
In order to constrain a class of models similar to RSB excess 2, we need to further reduce the instrumental noise or give a higher priority to the foregrounds.  An alternative strategy is to introduce priors on the RSB excess if necessary. In theory, if the uncertainty caused by the cosmic variance of a set of SH modes does not prevent us from constraining a particular RSB excess model with sufficient accuracy, then experiments measuring these modes can be used if the noise is low enough.

It should be noted that in real parameter estimation procedures we usually do not need to invert the covariance matrix, and therefore may not really need to reduce the sky degrees of freedom by making the universal SED assumption.
Furthermore, this assumption undoubtedly amplifies the contribution of the cosmic variance to the parameter uncertainty, since it essentially restricts all frequencies to carry the same information about the angular scale structure. In fact, more frequencies usually carry more information, even though the angular scale structures of galactic components may be highly correlated in frequency.
By understanding this, we can interpret the result with the universal SED assumption as a worst case, i.e. the upper limit of the uncertainties of real estimators.

\section{Discussion and conclusions}
\label{Sec: conclusion}
 {The origin of the RSB excess is a mystery.} Anisotropy measurements with 21cm arrays are a promising way to further constrain the origin.
In this work, we have developed a Fisher Forecast formalism for evaluating the potential for 21cm interferometric arrays to disentangle the anisotropic diffuse radio sky. 
In particular, we predict that a HERA-like array should be able to constrain the angular power spectrum (APS) of the RSB excess to a high degree even in the presence of multiple foreground components, with an uncertainty that depends on the true APS of the different components (Fig.~\ref{fig:RSB-corner-plots} shows toy model cases).

The formalism we developed assumes that each component of the radio sky can be well-approximated by a generalised parametric model.
The results of the Fisher analysis represent the uncertainties in the estimated parameters using a likelihood-based estimator that maximises a joint log-likelihood function of the form 
\begin{equation}
    \ln\mathcal{L} = \sum_s \, \ln \mathcal{L}^{(s)}\left(A^{(s)}, \alpha^{(s)}, \beta^{(s)}, \xi^{(s)}\Big\vert \, \mathbf{d}\right),
\end{equation}
for a given fiducial composition of the radio sky.
Ideally, we can maximise the likelihood function without using any informative priors, essentially putting all components of the sky on an equal footing. 
Extracting the RSB excess in this way (or perhaps also with the addition of ARCADE~2-based priors on some of the components) would in principle permit us to establish the existence of the RSB excess in a way that is independent of existing methods and results from monopole measurements. 
On the other hand, this component separation strategy will also be useful in deepening our understanding and tightening the constraints on all the remaining Galactic and extragalactic components, which would also facilitate other radio cosmology surveys that need to characterise and remove foreground emission.

If the existence of an RSB excess is confirmed, it would significantly change the interpretation of limits reported by 21cm surveys \citep[e.g.][]{Acharya2023,Fialkov2019}, which uses the primordial radiation background as a benchmark and relies on phenomenological foreground models for science extraction. 
If the potential background is emitted at several different redshifts, even before or during the Epoch of Reionization (EoR), then understanding the RSB becomes a prerequisite for understanding the 21cm signal. 

Discriminating between different theoretical explanations by measuring the angular power spectrum of {the RSB excess} may require a tighter constraint and thus a more elaborate experimental setup than simply detecting the presence of the RSB excess.
Our formalism provides two steps for uncertainty assessment to efficiently design observational strategies.
The first step is to select the appropriate set of spherical harmonic modes, as well as the frequency range and frequency resolution, by estimating the intrinsic uncertainty from cosmic variance (i.e. equation~\ref{eq: cosmic variance fisher}).
After determining the targeted modes and frequencies, the second step is to predict the uncertainty introduced by the experimental setup (equation~\ref{eq: data fisher 1}).
The advantage of radio interferometry is that since each baseline is most sensitive to the fluctuation mode at a particular scale, we can choose a particular combination of baselines to tailor the analysis. 
For example, if the frequency anticorrelation parameter $\xi$ for a theory is highly dependent on $\ell$, so we want to constrain this theoretical model only in a particular smaller range of $\ell$ where $\xi_\ell \simeq \xi'$ is approximately a constant, then we can down-weight the response to unwanted $\ell$ modes by filtering the baseline.

In the cosmic variance-limited (noise-free) case where all of the spherical harmonic modes are measured within a band between $20 \le \ell \le 90$ (or $40 \le \ell \le 70$; see Table~\ref{tab:frequency_SH_combinations}), we found that all of the foreground component parameters could be strongly constrained except for the (faint) Extragalactic Free-Free emission (Table~\ref{tab:cosmic_covariance}). This is in the case where there is no fiducial RSB excess model present, and no prior has been applied. This suggests that close-packed arrays of the kind used for 21cm surveys can usefully disentangle different anisotropic components of the radio sky based on angular and spectral clustering alone, i.e. without absolute spectrometry.

We also showed predictions for a HERA-like array to measure the angular power spectrum of the RSB excess for different toy models (results in Sec.~\ref{sec: hera results}). If the fiducial amplitude of the RSB excess component is of the same order as the dominant Galactic synchrotron component or larger, its amplitude and spectral/spatial power law indices can be constrained reasonably well within a typical observing season of $\sim 40$ nights or so, even if the spectral/spatial clustering properties are very similar to the Galactic synchrotron's. This is only the case if $\sim 10\%$ priors are applied to the non-RSB component parameters, as otherwise significant degeneracies arise. If the RSB excess is fainter than this at the target frequencies, however, it becomes difficult to disentangle from the foreground components based on angular and spectral clustering properties alone, and so either substantially more observing time, or stronger priors on the other components would be needed to permit its characterisation in terms of the parameters of our assumed APS model.

We reiterate that the universal SED approximation amplifies the contribution of the cosmic variance to the uncertainties of the parameter estimators.
The true uncertainty would theoretically be smaller than we have predicted, since a true estimator does not need to take the universal SED approximation.
Furthermore, if the `cosmic variance prediction' tells us that a selected set of sky modes are sufficient to constrain a theoretical model, then an experiment sensitive to these modes, provided it is sufficiently low-noise, should in principle be able to separate the components and constrain the model.

\section*{Acknowledgements}

We are grateful to I.~Browne, B.~Cyr, and P.~Wilkinson for useful discussions. This result is part of a project that has received funding from the European Research Council (ERC) under the European Union's Horizon 2020 research and innovation programme (Grant agreement No. 948764; ZZ, KAG, PB). 
We acknowledge use of the following software: 
{\tt matplotlib} \citep{matplotlib}, {\tt numpy} \citep{numpy}, {\tt corner} \citep{corner} and {\tt scipy} \citep{2020SciPy-NMeth}.  This work used the DiRAC@Durham facility managed by the Institute for Computational Cosmology on behalf of the STFC DiRAC HPC Facility (www.dirac.ac.uk). The equipment was funded by BEIS capital funding via STFC capital grants ST/P002293/1, ST/R002371/1 and ST/S002502/1, Durham University and STFC operations grant ST/R000832/1. DiRAC is part of the National e-Infrastructure.

\section*{Data Availability}

The Python code used to produce the results in this paper is available from \url{https://github.com/HydraRadio/Hydra}. Simulated data are available on request.



\balance

\bibliographystyle{mnras}
\bibliography{disentangling_radio_sky} 

@article{westerhout1951comparison,
  title={A comparison of the intensity distribution of radio-frequency radiation with a model of the galactic system},
  author={Westerhout, G and Oort, JH},
  journal={Bulletin of the Astronomical Institutes of the Netherlands, Vol. 11, p. 323},
  volume={11},
  pages={323},
  year={1951}
}

@article{dowell2018radio,
  title={The radio background below 100 MHz},
  author={Dowell, Jayce and Taylor, Greg B},
  journal={The Astrophysical Journal Letters},
  volume={858},
  number={1},
  pages={L9},
  year={2018},
  publisher={IOP Publishing}
}

@article{fixsen2011arcade,
  title={ARCADE 2 measurement of the absolute sky brightness at 3--90 GHz},
  author={Fixsen, DJ and Kogut, A and Levin, S and Limon, M and Lubin, P and Mirel, P and Seiffert, M and Wollack, E and Villela, T and others},
  journal={The Astrophysical Journal},
  volume={734},
  number={1},
  pages={5},
  year={2011},
  publisher={IOP Publishing}
}

@article{singal2023second,
  title={The Second Radio Synchrotron Background Workshop: Conference Summary and Report},
  author={Singal, J and Fornengo, N and Regis, M and Bernardi, G and Bordenave, D and Branchini, E and Cappelluti, N and Caputo, A and Carucci, IP and Chluba, J and others},
  journal={Publications of the Astronomical Society of the Pacific},
  volume={135},
  number={1045},
  pages={036001},
  year={2023},
  publisher={IOP Publishing}
}

@article{cyr2023cosmic,
  title={A cosmic string solution to the radio synchrotron background},
  author={Cyr, Bryce and Chluba, Jens and Acharya, Sandeep Kumar},
  journal={arXiv preprint arXiv:2308.03512},
  year={2023}
}

@article{vernstrom2014deep,
  title={Deep 3 GHz number counts from a P (D) fluctuation analysis},
  author={Vernstrom, T and Scott, Douglas and Wall, JV and Condon, JJ and Cotton, WD and Fomalont, EB and Kellermann, KI and Miller, N and Perley, RA},
  journal={Monthly Notices of the Royal Astronomical Society},
  volume={440},
  number={3},
  pages={2791--2809},
  year={2014},
  publisher={Oxford University Press}
}

@article{condon2012resolving,
  title={Resolving the radio source background: deeper understanding through confusion},
  author={Condon, JJ and Cotton, WD and Fomalont, EB and Kellermann, KI and Miller, N and Perley, RA and Scott, D and Vernstrom, T and Wall, JV},
  journal={The Astrophysical Journal},
  volume={758},
  number={1},
  pages={23},
  year={2012},
  publisher={IOP Publishing}
}

@article{hardcastle2021contribution,
  title={The contribution of discrete sources to the sky temperature at 144 MHz},
  author={Hardcastle, MJ and Shimwell, TW and Tasse, C and Best, PN and Drabent, A and Jarvis, MJ and Prandoni, ISABELLA and R{\"o}ttgering, HJA and Sabater, J and Schwarz, DJ},
  journal={Astronomy \& Astrophysics},
  volume={648},
  pages={A10},
  year={2021},
  publisher={EDP Sciences}
}

@article{cline2013cosmological,
  title={Cosmological origin of anomalous radio background},
  author={Cline, James M and Vincent, Aaron C},
  journal={Journal of Cosmology and Astroparticle Physics},
  volume={2013},
  number={02},
  pages={011},
  year={2013},
  publisher={IOP Publishing}
}

@article{ewall2018modeling,
  title={Modeling the radio background from the first black holes at cosmic dawn: implications for the 21 cm absorption amplitude},
  author={Ewall-Wice, A and Chang, T-C and Lazio, J and Dor{\'e}, O and Seiffert, M and Monsalve, RA},
  journal={The Astrophysical Journal},
  volume={868},
  number={1},
  pages={63},
  year={2018},
  publisher={IOP Publishing}
}

@article{mittal2022background,
  title={Background of radio photons from primordial black holes},
  author={Mittal, Shikhar and Kulkarni, Girish},
  journal={Monthly Notices of the Royal Astronomical Society},
  volume={510},
  number={4},
  pages={4992--4997},
  year={2022},
  publisher={Oxford University Press}
}

@article{fortes2019some,
  title={Some implications of the leptonic annihilation of dark matter: possible galactic radio emission signatures and the excess radio flux of extragalactic origin},
  author={Fortes, Elaine CFS and Miranda, Oswaldo D and Stecker, Floyd W and Wuensche, Carlos A},
  journal={Journal of Cosmology and Astroparticle Physics},
  volume={2019},
  number={11},
  pages={047},
  year={2019},
  publisher={IOP Publishing}
}

@article{offringa2022measurement,
  title={Measurement of the anisotropy power spectrum of the radio synchrotron background},
  author={Offringa, AR and Singal, J and Heston, S and Horiuchi, S and Lucero, DM},
  journal={Monthly Notices of the Royal Astronomical Society},
  volume={509},
  number={1},
  pages={114--121},
  year={2022},
  publisher={Oxford University Press}
}

@article{cowie2023diffuse,
  title={Diffuse sources, clustering and the excess anisotropy of the radio synchrotron background},
  author={Cowie, FJ and Offringa, AR and Gehlot, BK and Singal, J and Heston, S and Horiuchi, S and Lucero, DM},
  journal={Monthly Notices of the Royal Astronomical Society},
  pages={stad1671},
  year={2023},
  publisher={Oxford University Press}
}

@article{biermann2014cosmic,
  title={Cosmic backgrounds due to the formation of the first generation of supermassive black holes},
  author={Biermann, Peter L and Nath, Biman B and Caramete, Lauren{\c{t}}iu I and Harms, Benjamin C and Stanev, Todor and Tjus, Julia Becker},
  journal={Monthly Notices of the Royal Astronomical Society},
  volume={441},
  number={2},
  pages={1147--1156},
  year={2014},
  publisher={Oxford University Press}
}

@article{santos2005multifrequency,
  title={Multifrequency analysis of 21 centimeter fluctuations from the era of reionization},
  author={Santos, M{\'a}rio G and Cooray, Asantha and Knox, Lloyd},
  journal={The Astrophysical Journal},
  volume={625},
  number={2},
  pages={575},
  year={2005},
  publisher={IOP Publishing}
}

@article{de2008model,
  title={A model of diffuse Galactic radio emission from 10 MHz to 100 GHz},
  author={de Oliveira-Costa, Ang{\'e}lica and Tegmark, Max and Gaensler, BM and Jonas, Justin and Landecker, TL and Reich, Patricia},
  journal={Monthly Notices of the Royal Astronomical Society},
  volume={388},
  number={1},
  pages={247--260},
  year={2008},
  publisher={The Royal Astronomical Society}
}

@article{deboer2017hydrogen,
  title={Hydrogen epoch of reionization array (HERA)},
  author={DeBoer, David R and Parsons, Aaron R and Aguirre, James E and Alexander, Paul and Ali, Zaki S and Beardsley, Adam P and Bernardi, Gianni and Bowman, Judd D and Bradley, Richard F and Carilli, Chris L and others},
  journal={Publications of the Astronomical Society of the Pacific},
  volume={129},
  number={974},
  pages={045001},
  year={2017},
  publisher={IOP Publishing}
}

@ARTICLE{Acharya2023,
       author = {{Acharya}, Sandeep Kumar and {Cyr}, Bryce and {Chluba}, Jens},
        title = "{The role of soft photon injection and heating in 21 cm cosmology}",
      journal = {\mnras},
         year = 2023,
        month = aug,
       volume = {523},
       number = {2},
        pages = {1908-1918},
          doi = {10.1093/mnras/stad1540},
archivePrefix = {arXiv},
       eprint = {2303.17311},
 primaryClass = {astro-ph.CO},
       adsurl = {https://ui.adsabs.harvard.edu/abs/2023MNRAS.523.1908A}
}

@article{Fialkov2019,
    author = {Fialkov, Anastasia and Barkana, Rennan},
    title = "{Signature of excess radio background in the 21-cm global signal and power spectrum}",
    journal = {Monthly Notices of the Royal Astronomical Society},
    volume = {486},
    number = {2},
    pages = {1763-1773},
    year = {2019},
    month = {03},
    issn = {0035-8711},
    doi = {10.1093/mnras/stz873},
    url = {https://doi.org/10.1093/mnras/stz873},
    eprint = {https://academic.oup.com/mnras/article-pdf/486/2/1763/28484631/stz873.pdf},
}

@misc{kittiwisit2023matvis,
      title={matvis: A matrix-based visibility simulator for fast forward modelling of many-element 21 cm arrays}, 
      author={Piyanat Kittiwisit and Steven G. Murray and Hugh Garsden and Philip Bull and Christopher Cain and Aaron R. Parsons and Jackson Sipple and Zara Abdurashidova and Tyrone Adams and James E. Aguirre and Paul Alexander and Zaki S. Ali and Rushelle Baartman and Yanga Balfour and Adam P. Beardsley and Lindsay M. Berkhout and Gianni Bernardi and Tashalee S. Billings and Judd D. Bowman and Richard F. Bradley and Jacob Burba and Steven Carey and Chris L. Carilli and Kai-Feng Chen and Carina Cheng and Samir Choudhuri and David R. DeBoer and Eloy de Lera Acedo and Matt Dexter and Joshua S. Dillon and Scott Dynes and Nico Eksteen and John Ely and Aaron Ewall-Wice and Nicolas Fagnoni and Randall Fritz and Steven R. Furlanetto and Kingsley Gale-Sides and Bharat Kumar Gehlot and Abhik Ghosh and Brian Glendenning and Adelie Gorce and Deepthi Gorthi and Bradley Greig and Jasper Grobbelaar and Ziyaad Halday and Bryna J. Hazelton and Jacqueline N. Hewitt and Jack Hickish and Tian Huang and Daniel C. Jacobs and Alec Josaitis and Austin Julius and MacCalvin Kariseb and Nicholas S. Kern and Joshua Kerrigan and Honggeun Kim and Saul A. Kohn and Matthew Kolopanis and Adam Lanman and Paul La Plante and Adrian Liu and Anita Loots and Yin-Zhe Ma and David H. E. MacMahon and Lourence Malan and Cresshim Malgas and Keith Malgas and Bradley Marero and Zachary E. Martinot and Andrei Mesinger and Mathakane Molewa and Miguel F. Morales and Tshegofalang Mosiane and Abraham R. Neben and Bojan Nikolic and Chuneeta Devi Nunhokee and Hans Nuwegeld and Robert Pascua and Nipanjana Patra and Samantha Pieterse and Yuxiang Qin and Eleanor Rath and Nima Razavi-Ghods and Daniel Riley and James Robnett and Kathryn Rosie and Mario G. Santos and Peter Sims and Saurabh Singh and Dara Storer and Hilton Swarts and Jianrong Tan and Nithyanandan Thyagarajan and Pieter van Wyngaarden and Peter K. G. Williams and Zhilei Xu and Haoxuan Zheng},
      year={2023},
      eprint={2312.09763},
      archivePrefix={arXiv},
      primaryClass={astro-ph.IM}
}

@ARTICLE{matplotlib,
  author={J. D. {Hunter}},
  journal={Computing in Science   Engineering}, 
  title={Matplotlib: A 2D Graphics Environment}, 
  year={2007},
  volume={9},
  number={3},
  pages={90-95},
}

@ARTICLE{numpy,
  author={S. {van der Walt} and S. C. {Colbert} and G. {Varoquaux}},
  journal={Computing in Science   Engineering}, 
  title={The NumPy Array: A Structure for Efficient Numerical Computation}, 
  year={2011},
  volume={13},
  number={2},
  pages={22-30},
}

@ARTICLE{2021Padovani,
       author = {{Padovani}, Marco and {Bracco}, Andrea and {Jeli{\'c}}, Vibor and {Galli}, Daniele and {Bellomi}, Elena},
        title = "{Spectral index of synchrotron emission: insights from the diffuse and magnetised interstellar medium}",
      journal = {\aap},
         year = 2021,
        month = jun,
       volume = {651},
          eid = {A116},
        pages = {A116},
          doi = {10.1051/0004-6361/202140799},
archivePrefix = {arXiv},
       eprint = {2106.10929},
 primaryClass = {astro-ph.HE},
       adsurl = {https://ui.adsabs.harvard.edu/abs/2021A&A...651A.116P}
}

@ARTICLE{2020SciPy-NMeth,
       author = {{Virtanen}, Pauli and {Gommers}, Ralf and {Oliphant},
         Travis E. and {Haberland}, Matt and {Reddy}, Tyler and
         {Cournapeau}, David and {Burovski}, Evgeni and {Peterson}, Pearu
         and {Weckesser}, Warren and {Bright}, Jonathan and {van der Walt},
         St{\'e}fan J.  and {Brett}, Matthew and {Wilson}, Joshua and
         {Jarrod Millman}, K.  and {Mayorov}, Nikolay and {Nelson}, Andrew
         R.~J. and {Jones}, Eric and {Kern}, Robert and {Larson}, Eric and
         {Carey}, CJ and {Polat}, {\.I}lhan and {Feng}, Yu and {Moore},
         Eric W. and {Vand erPlas}, Jake and {Laxalde}, Denis and
         {Perktold}, Josef and {Cimrman}, Robert and {Henriksen}, Ian and
         {Quintero}, E.~A. and {Harris}, Charles R and {Archibald}, Anne M.
         and {Ribeiro}, Ant{\^o}nio H. and {Pedregosa}, Fabian and
         {van Mulbregt}, Paul and {Contributors}, SciPy 1. 0},
        title = "{SciPy 1.0: Fundamental Algorithms for Scientific
                  Computing in Python}",
      journal = {Nature Methods},
      year = "2020",
      volume={17},
      pages={261--272},
      adsurl = {https://rdcu.be/b08Wh},
      doi = {https://doi.org/10.1038/s41592-019-0686-2},
}

@article{corner,
      doi = {10.21105/joss.00024},
      url = {https://doi.org/10.21105/joss.00024},
      year  = {2016},
      month = {jun},
      publisher = {The Open Journal},
      volume = {1},
      number = {2},
      pages = {24},
      author = {Daniel Foreman-Mackey},
      title = {corner.py: Scatterplot matrices in Python},
      journal = {The Journal of Open Source Software}
    }

@article{todarello2023,
    author = {Todarello, Elisa and Regis, Marco and Bianchini, Federico and Singal, Jack and Branchini, Enzo and Cowie, Fraser J and Heston, Sean and Horiuchi, Shunsaku and Lucero, Danielle and Offringa, Andre},
    title = "{Constraints on the origin of the radio synchrotron background via angular correlations}",
    journal = {Monthly Notices of the Royal Astronomical Society},
    pages = {stae876},
    year = {2024},
    month = {03},
    issn = {0035-8711},
    doi = {10.1093/mnras/stae876},
    url = {https://doi.org/10.1093/mnras/stae876},
    eprint = {https://academic.oup.com/mnras/advance-article-pdf/doi/10.1093/mnras/stae876/57098007/stae876.pdf},
}






\bsp	
\label{lastpage}
\end{document}